\documentstyle [11pt,epsfig] {article}

\def\ds{\displaystyle}
\def\Tr{\mbox{Tr}\,}

\begin{document}

\noindent ULM--TP/96-4 \\
November 1996
\vspace{2.4cm}

\centerline{\huge Semiclassical Transition from an Elliptical}
\vspace{0.3cm}
\centerline{\huge to an Oval Billiard}
\vspace{1.3cm}
\centerline{\Large Martin Sieber$^{1,2}$}
\vspace{0.5 cm}

\noindent $^1$ Division de Physique Th\'eorique\footnotemark,
Institut de Physique Nucl\'eaire, F-91406 Orsay Cedex, France \\
\noindent $^2$ Abteilung Theoretische Physik\footnotemark, 
Universit\"at Ulm, D-89069 Ulm, Germany

\vspace{1.8 cm}
\centerline{\bf Abstract}
\vspace{0.5cm}

Semiclassical approximations often involve the use of
stationary phase approximations.
This method can be applied when $\hbar$ is small in comparison
to relevant actions or action differences in the corresponding
classical system. 
In many situations, however, action differences can be
arbitrarily small and then uniform approximations are more
appropriate. In the present paper we examine different uniform
approximations for describing the spectra of integrable systems
and systems with mixed phase space.
This is done on the example of two billiard systems, an
elliptical billiard and a deformation of it, an oval billiard. 
We derive a trace formula for the ellipse which involves
a uniform approximation for the Maslov phases near the
separatrix, and a uniform approximation for tori of 
periodic orbits close to a bifurcation. We then 
examine how the trace formula is modified when the
ellipse is deformed into an oval. This involves uniform
approximations for the break-up of tori and uniform
approximations for bifurcations of periodic orbits.
Relations between different uniform approximations are
discussed.

\vspace{2.0cm}
\begin{math}
\footnotetext[1]{Unit\'e de Recherche des Universit\'es Paris
XI et Paris VI associ\'ee au CNRS}
\footnotetext[2]{Present address}
\end{math}

\noindent PACS numbers: \\
\noindent 03.65.Ge ~ Solutions of wave equations: bound states. \\
\noindent 03.65.Sq ~ Semiclassical theories and applications. \\
\noindent 05.45.+b ~ Theory and models of chaotic systems.

\vspace{0.3cm}
\noindent{\it Submitted to Journal of Physics A}

\newpage

\section{Introduction}

Semiclassical trace formulas are important tools for
analyzing spectra of quantum systems. They have
found a wide range of applications in recent years
(see for example \cite{Cha92,Cha93,CC95}).
Trace formulas have been derived for integrable
and chaotic systems when periodic orbits lie on tori
in phase space or are isolated \cite{Gut71,BB72,BT76,BT77a,Gut90}.
Families of orbits in systems with more general symmetries
have also been treated \cite{CL91,CL92}.

Most systems, however, are neither integrable nor chaotic,
but have a phase space which is intricately divided into
regular and chaotic regions. In these systems there exist
classical structures on all scales, and neighbouring periodic
orbits can have arbitrarily small action differences.
For the semiclassical approximation this has the consequence
that in many cases stationary phase approximations cannot be
applied since they rely on the fact that action differences of
periodic orbits are large in comparison to $\hbar$. Instead
one has to use approximations which are uniformly valid in two
small parameters, $\hbar$ and $\varepsilon$, where 
$\varepsilon$ describes the separation of neighbouring
periodic orbits.

A typical situation where small action differences of
periodic orbits occur is near bifurcations of periodic orbits. 
Bifurcations are a ubiquitous phenomenon in mixed systems.
They occur any time the stability angle of a stable
orbit is a rational multiple of $2 \pi$. If, for example,
one changes an external parameter of a system by an
arbitrarily small but finite amount (or the energy in a
generic system), then in general an infinite number of
bifurcations occur, most of them for long periodic
orbits.

Bifurcations occur in different forms, but the number
of generic forms in two-dimensional systems is finite.
They were classified by Meyer \cite{Mey70} and Bruno
\cite{Brj70,Bru72}. Their form depends on the repetition
number $m$ of an orbit that bifurcates. The semiclassical
treatment of these generic bifurcations was investigated by
Ozorio de Almeida and Hannay \cite{OH87}. They derived 
contributions to the trace formula from periodic orbits
near bifurcations in terms of standard diffraction
catastrophe integrals. These approximations have 
for example been applied for
treating tangent bifurcations and pitchfork
bifurcations \cite{KHD93,AE94}.
For the largest class of bifurcations
with $m>4$ the results of Ozorio de Almeida and Hannay
were extended in \cite{Sie96} by including higher order
terms in the normal form expansion for the
bifurcation. In this way a uniform approximation
was obtained which interpolates over the regime
from the bifurcation up to regions where Gutzwiller's
approximation can be applied.

Another situation in which small action differences occur
is in the quasi-integrable regime, i.\,e.\ for small
perturbations of integrable systems. Due to the perturbation
all rational tori of the integrable system break up into
a finite number of periodic orbits which have small action
differences when the perturbation is small. Ozorio de Almeida
derived a uniform approximation which is valid if the splitting
of the orbits is small \cite{Ozo86}. A formula for the break-up of
families of orbits due to more general
symmetries was derived by Creagh \cite{Cre96}.
Tomsovic, Grinberg and Ullmo extended the result of
Ozorio de Almeida and obtained a uniform approximation
which interpolates between the torus contribution and
Gutzwiller's approximation \cite{TGU95,UGT96}.

Small action differences of neighbouring periodic
orbits can occur also in integrable systems, since
bifurcations of periodic orbits can occur also there.
These bifurcations do not belong to the class of
generic bifurcations, since they result
in the appearance of whole new tori of periodic orbits.
A uniform approximation for the contributions of
tori that result from the bifurcation of a stable
orbit was derived by Berry and Tabor \cite{BT76} 
and Richens \cite{Ric82}.

The present paper contains an investigation and 
discussion of the various concepts which are
mentioned above. This is done on the example
of two billiard systems, an ellipse and a
deformation of it, an oval. 
The classical motion in an ellipse is integrable
and has several characteristic properties.
There is a separatrix which separates two kinds
of motion, around the two foci and between the
two foci. 
Furthermore, in addition to the usual tori
of periodic orbits there are also two isolated
periodic orbits in the ellipse (and their repetitions),
one stable and one unstable. When the eccentricity
of the ellipse is increased, then new tori of
periodic orbits arise through bifurcations of 
the stable orbit and its repetitions. All these
classical properties have an influence on the
semiclassical approximation. After a brief review
of the classical and quantum mechanics in the
ellipse we discuss the semiclassical EBK-quantization
for the ellipse. Near the separatrix this quantization
condition has to be modified by a (non-integer)
uniform approximation for the Maslov index, since
this index is discontinuous at the separatrix
which leads to ambiguities and inaccuracies in the
semiclassical quantization \cite{AA87a,AA87b}.
From this modified EBK-quantization condition
we derive a trace formula for the ellipse.
We show in detail, how the contributions of regular
tori of orbits, of tori near bifurcations, of the
isolated periodic orbits and the area and perimeter
contributions can be obtained from the modified
EBK-quantization conditions. We close this section
with a numerical examination of the trace formula.

We then examine an oval-shaped billiard which
can be considered as a deformation of
an ellipse. We investigate how
the semiclassical approximation for the
ellipse is changed when the ellipse is deformed
into the oval. Due to this perturbation
all tori of periodic orbits
break up into a finite number of orbits. 
When the tori are not close to a bifurcation
then the uniform approximation for the 
break-up of tori can be applied. For 
tori near bifurcations one has to 
use a different approximation.
It will be shown that the (slightly modified)
uniform approximations for generic bifurcations
describe also the break-up of the tori near 
bifurcations. We present a detailed
numerical examination of semiclassical
approximations in the oval, and we
discuss the relations between
different uniform approximations that
are used in this paper.

\section{The elliptical billiard}

We discuss in this section various properties of the elliptical
billiard. This is an integrable billiard system whose boundary
is defined by
\begin{equation} \label{e1}
\frac{x^2}{a^2} + \frac{y^2}{b^2} = 1 \; ,
\end{equation}
where $a$ and $b<a$ are the lengths of the semi-major and 
semi-minor axis, respectively. The ellipse has two focal
points with coordinates $(\pm c,0)$, where
$c = \sqrt{a^2 - b^2}$, and its eccentricity 
is defined by $e = c/a$. The billiard area is
$A = \pi a b$ and its perimeter is $L = 4 a \mbox{E}(e)$
where $\mbox{E}(\kappa)$ is the complete elliptic integral
of the second kind \cite{GR80} (see Eq.\ (\ref{e11}) below).

The classical properties in the elliptical billiard have been
investigated under different view points. This includes the
treatment of the action-angle variables \cite{KR60}, 
the caustics of the classical motion \cite{KR60,Sin76}, 
Poncelet's theorem \cite{CF88,CCS93b}, the billiard map
\cite{Ber81,Tab94}, and the periodic orbits 
\cite{CCS93b,ONO90,RWKK95}.
A treatment of the Schr\"odinger equation for the elliptical
billiard can be found in \cite{MF53}. In the following we briefly
review classical and quantum properties of the elliptical billiard
because they are used for the semiclassical approximation.
We then discuss the EBK-quantization for the billiard and a
uniform extension of it, and we derive a trace formula in
terms of the periodic orbits of the system.

\subsection{Classical mechanics of the elliptical billiard}
\label{secclass}

The classical motion of a particle in an elliptical billiard is
conveniently described in elliptical coordinates
\begin{eqnarray} \label{e2}
x &=& c \, \cosh u  \, \cos v  \nonumber \\
y &=& c \, \sinh u  \, \sin v \; ,
\end{eqnarray}
where $u$ is restricted to $0 \le u \le \mbox{arctanh}(b/a)$ and 
$v$ is a periodic coordinate with period $2 \pi$. 
In terms of these coordinates the Lagrangian has the form
(with mass $m = 1/2$)
\begin{equation} \label{e3}
L = \frac{c^2}{4} (\cosh^2 u - \cos^2 v)(\dot{u}^2 + \dot{v}^2) \; ,
\end{equation}
the canonical momenta are given by
\begin{eqnarray} \label{e4}
p_u &=& \frac{\partial L}{\partial \dot{u}}
    = \frac{c^2}{2} \dot{u} \, (\cosh^2 u - \cos^2 v)  \nonumber \\
p_v &=& \frac{\partial L}{\partial \dot{v}}
    = \frac{c^2}{2} \dot{v} \, (\cosh^2 u - \cos^2 v) \; ,
\end{eqnarray}
and the Hamiltonian is
\begin{equation} \label{e5}
H = \frac{p_u^2 + p_v^2}{c^2 (\cosh^2 u - \cos^2 v)} \; .
\end{equation}

The two conserved quantities of the system are the energy and
the product $L_1 L_2$ of the two angular momenta about the two
foci
\begin{equation} \label{e6}
L_1 L_2 = \frac{p_v^2 \sinh^2 u - p_u^2 \sin^2 v}{\cosh^2 u - 
\cos^2 v} \; .
\end{equation}
It is more convenient to use instead of $L_1 L_2$ another
conserved quantity which is energy independent and is determined
by the geometrical properties of a trajectory only. This is 
$\alpha = L_1 L_2 / E$ whose values are restricted to the range
$b^2 \ge \alpha \ge -c^2$. The upper limit $\alpha = b^2$
corresponds to the motion along the boundary and the lower limit 
$\alpha = -c^2$ corresponds to the motion along the minor axis.

In terms of $E$ and $\alpha$ the canonical momenta are given by
\begin{eqnarray} \label{e7}
p_u^2 &=& E (c^2 \sinh^2 u - \alpha) \nonumber \\
p_v^2 &=& E (c^2 \sin^2 v  + \alpha) \; .
\end{eqnarray}

There are two different kinds of motion in the ellipse depending
on the sign of $\alpha$.
For $0 < \alpha \leq b^2$ the trajectories have a caustic in form 
of a confocal ellipse with semi-minor axis $b' = \sqrt{\alpha}$.
The motion goes around the two focal points and is composed
of a libration in the coordinate $u$ and a rotation in the
coordinate $v$. 
For $-c^2 \leq \alpha < 0$ the caustic of the motion is a confocal
hyperbola with semi-conjugate axis $b' = \sqrt{-\alpha}$
and semi-transverse axis $a' = \sqrt{c^2 + \alpha}$. The motion
is composed of a libration in the coordinate $u$ and a libration
in the coordinate $v$, and the trajectories cross the $x$-axis 
always between the two focal points.
Both kinds of motions are separated by a separatrix which consists
of orbits with $\alpha = 0$ that go through the focal points of
the ellipse.

The periodic orbits of the elliptical billiard can be determined
by introducing action-angle variables. The actions are given by
\begin{eqnarray} \label{e8}
I_u &=& \frac{1}{2 \pi} \oint \! p_u \, \mbox{d} u
= \frac{\sqrt{E}}{\pi} \int_{u_0}^{u_1} \!
\mbox{d} u \, \sqrt{c^2 \sinh^2 u - \alpha}  \nonumber \\
I_v &=& \frac{1}{2 \pi} \oint \! p_v \, \mbox{d} v 
= \frac{2 \sqrt{E}}{\pi} \int_{v_0}^{v_1} \! 
\mbox{d} v \, \sqrt{c^2 \sin^2 v + \alpha} \; ,
\end{eqnarray} 
where
\begin{equation} \label{e9}
\begin{array}{lllll}                      \ds
\mbox{for} \; \alpha > 0:               \hspace{2ex} & \ds
u_0 = \mbox{arcsinh} \frac{b'}{c} \, ,  \hspace{1ex} & \ds
u_1 = \mbox{arcsinh} \frac{b}{c}  \, ,  \hspace{1ex} & \ds
v_0 = 0                           \, ,  \hspace{1ex} & \ds
v_1 = \frac{\pi}{2}               \, ,  \\[6pt] \ds
\mbox{for} \; \alpha < 0:               \hspace{2ex} & \ds
u_0 = 0                           \, ,  \hspace{1ex} & \ds
u_1 = \mbox{arcsinh} \frac{b}{c}  \, ,  \hspace{1ex} & \ds
v_0 = \arcsin \frac{b'}{c}        \, ,  \hspace{1ex} & \ds
v_1 = \frac{\pi}{2}               \, ,
\end{array}
\end{equation}
and $b' = \sqrt{|\alpha|}$. We choose here a definition of $I_u$ 
which is actually the action for only half a cycle if $\alpha < 0$.
The reason is that otherwise $I_u$ is discontinuous if $\alpha$
changes sign. We remark that the actions in a half-ellipse which
is cut along the $x$-axis have the values given by Eqs.\ (\ref{e8})
and (\ref{e9}).

The integrations in (\ref{e8}) yield
\begin{eqnarray} \label{e10}
I_u &=& \left\{ \begin{array}{ll} \ds
\frac{\sqrt{E}}{\pi} \left[ \frac{a}{b} \sqrt{b^2 - \alpha}
- \frac{c}{\kappa} \, \mbox{E}(\arcsin 
\sqrt{\frac{b^2 - \alpha}{b^2}}, \kappa) \right]
& \mbox{for} \; \alpha > 0 \\ [6pt] \ds
\frac{\sqrt{E}}{\pi} \left[ \frac{ab}{\sqrt{b^2 - \alpha}}
- \frac{\alpha}{c} \, \mbox{F}(\arcsin 
\sqrt{\frac{b^2}{b^2 - \alpha}}, \frac{1}{\kappa})
- c \, \mbox{E}(\arcsin \sqrt{\frac{b^2}{b^2 - \alpha}},
\frac{1}{\kappa}) \right]
& \mbox{for} \; \alpha < 0 
\end{array} \right. \nonumber \\
I_v &=& \left\{ \begin{array}{ll} \ds
\frac{2 \sqrt{E}}{\pi} \, \frac{c}{\kappa} \, 
\mbox{E}(\frac{\pi}{2},\kappa)
& \phantom{blablablablablablablablablablabla} 
\mbox{for} \; \alpha > 0 \\ [6pt] \ds 
\frac{2 \sqrt{E}}{\pi} \left[ 
\frac{\alpha}{c} \, \mbox{F}(\frac{\pi}{2}, \frac{1}{\kappa})
+ c \, \mbox{E}(\frac{\pi}{2},\frac{1}{\kappa}) \right]
& \phantom{blablablablablablablablablablabla} 
\mbox{for} \; \alpha < 0 \; ,
\end{array} \right.
\end{eqnarray} 
where $\kappa = c/a' = c/\sqrt{c^2 + \alpha}$, and the functions
\begin{equation} \label{e11}
\mbox{F}(\varphi,\kappa) =  \int_0^\varphi \!
\frac{\mbox{d} \alpha}{\sqrt{1 - \kappa^2 \sin^2 \alpha}} 
\; , \hspace{5ex} \; 
\mbox{E}(\varphi,\kappa) =  \int_0^\varphi \! 
\sqrt{1 - \kappa^2 \sin^2 \alpha} \, \mbox{d} \alpha \; ,
\end{equation}
are the elliptic integrals of the first and second kind \cite{GR80}.
The quantity $\kappa$ is called the modulus of the elliptic integrals.
The corresponding complete elliptic integrals are denoted by
$\mbox{E}(\kappa) = \mbox{E}(\pi/2,\kappa)$ and
$\mbox{K}(\kappa) = \mbox{F}(\pi/2,\kappa)$.

The tori of periodic orbits of the elliptical billiard are determined
by the condition that the quotient of the angular frequencies
$\omega_u$ and $\omega_v$ is rational:
\begin{equation} \label{e12}
\frac{\omega_u}{\omega_v}
= \frac{\left. \frac{\partial E}{\partial I_u} \right|_{I_v} }{
        \left. \frac{\partial E}{\partial I_v} \right|_{I_u} }
= -\left. \frac{\partial I_v}{\partial I_u} \right|_E
= -\frac{\left. \frac{\partial I_v}{\partial \alpha} \right|_E }{
         \left. \frac{\partial I_u}{\partial \alpha} \right|_E }
= \frac{n}{m} \; ,
\end{equation}
and with 
\begin{eqnarray} \label{e13}
\left. \frac{\partial I_u}{\partial \alpha} \right|_E 
&=& \left\{ \begin{array}{ll} \ds
- \frac{\sqrt{E} \kappa}{2 \pi c} \,
\mbox{F}\left( 
\arcsin \sqrt{\frac{b^2 - \alpha}{b^2}},\kappa\right)
\hspace{3ex} & \mbox{for} \; \alpha > 0 \\[8pt] \ds
- \frac{\sqrt{E}}{2 \pi c} \,
\mbox{F}\left(
\arcsin \sqrt{\frac{b^2}{b^2 - \alpha}},\frac{1}{\kappa}\right)
\hspace{3ex} & \mbox{for} \; \alpha < 0 
\end{array} \right. \nonumber \\
\left. \frac{\partial I_v}{\partial \alpha} \right|_E 
&=& \left\{ \begin{array}{ll} \ds
\frac{\kappa \sqrt{E}}{\pi c} \, \mbox{F}\left(
\frac{\pi}{2},\kappa\right)
& \hspace{19ex} \mbox{for} \; \alpha > 0 \\[8pt] \ds
\frac{\sqrt{E}}{\pi c} \, \mbox{F}\left(
\frac{\pi}{2},\frac{1}{\kappa}\right)
& \hspace{19ex} \mbox{for} \; \alpha < 0 \; ,
\end{array} \right.
\end{eqnarray} 
one obtains the conditions for periodic orbits.
We will consider in the following the cases $\alpha > 0$ and
$\alpha < 0$ separately.

\

\noindent \underline{The case $\alpha > 0$:}

\vspace{2mm} \noindent
The condition for periodic orbits is 
\begin{equation} \label{e14}
\mbox{F}(\arcsin \sqrt{\frac{b^2 - \alpha}{b^2}},\kappa) =
\frac{2 m}{n} \, \mbox{F}(\frac{\pi}{2},\kappa) =
\frac{2 m}{n} \, \mbox{K}(\kappa) \; .
\end{equation}
This equation has a solution for all
$n=3,4,\dots$ and $1 \le m < n/2$, and the integers $n$ and
$m$ are the number of reflections of an orbit and its rotation
number, respectively. Eq.\,(\ref{e14}) determines the
value of $\alpha$ for a family of periodic orbits.

We briefly discuss the case $m=n/2$. For this case
there exist also periodic orbits which are the orbit along
the major axis and its repetitions. However, these orbits are
isolated and do not appear in families. This is reflected
by the fact that for $m=n/2$, there exists no solution of
Eq.\ (\ref{e14}) (although it does exist for Eq.\,(\ref{e12})).
In the limit $\alpha \rightarrow 0$,
the first argument of the elliptic integral on the right-hand
side of Eq.\,(\ref{e14}) approaches $\pi/2$, but the integral
itself is divergent in this limit since $\kappa \rightarrow 1$.
The correct limiting behaviour can be obtained from the
relation \cite{AS65}
\begin{equation} \label{e15}
\mbox{F}(\varphi,\kappa) + \mbox{F}(\psi,\kappa) = 
\mbox{F}(\frac{\pi}{2},\kappa)  \; \; \; \; \;
\mbox{if} \; \; \; \; \;  
\sqrt{1 - \kappa^2} \, \tan \varphi \, \tan \psi = 1 \; .  
\end{equation}
From this relation follows that
\begin{equation} \label{e16}
\lim_{\alpha \rightarrow 0} \left[
\mbox{F}(\frac{\pi}{2},\kappa) -
\mbox{F}(\arcsin \sqrt{\frac{b^2 - \alpha}{b^2}},\kappa) \right]
= \mbox{F}(\arctan \frac{c}{b},1) 
= \log \frac{a+b}{c} \; .
\end{equation}

A different way of writing condition (\ref{e14}) is obtained by
inverting the equation which leads to
\begin{equation} \label{e17}
\sqrt{\frac{b^2 - \alpha}{b^2}} 
= \mbox{sn}(\frac{2m}{n} \mbox{K}) \; ,
\end{equation}
where sn($u$) is a Jacobian elliptic function. In the above
notation, the modulus $\kappa$, which acts as an independent
variable, is omitted.

The lengths of the periodic orbits are given by
\begin{eqnarray} \label{e18}
l_{n,m} &=& \frac{2 \pi}{\sqrt{E}} (n I_u + m I_v) \nonumber \\
&=& \frac{2 n a}{b} \sqrt{b^2 - \alpha} - \frac{2 n c}{\kappa} \,
\mbox{E}(\arcsin \sqrt{\frac{b^2 - \alpha}{b^2}},\kappa)
+ \frac{4 m c}{\kappa} \, \mbox{E}(\frac{\pi}{2},\kappa) \nonumber \\
&=& \frac{2 n a}{b} \sqrt{b^2 - \alpha} - \frac{2 n c}{\kappa} \, 
\mbox{Z}(\frac{2m}{n} \mbox{K}) \; ,
\end{eqnarray}
where Eqs.\,(\ref{e10}) and (\ref{e14}) have been used and
Z($u$) is Jacobi's zeta function
\begin{equation} \label{e19}
\mbox{Z}(u) = \mbox{E}(\varphi,\kappa) - \mbox{F}(\varphi,\kappa)
\, \frac{\mbox{E}(\pi/2,\kappa)}{\mbox{F}(\pi/2,\kappa)} \; .
\end{equation}
Here $u=\mbox{F}(\varphi,\kappa)$ and the dependence of $Z(u)$ on
the modulus $\kappa$ is again omitted in the notation.

Addition theorems for Jacobian elliptic functions
and Jacobi's zeta function can be used to obtain algebraic
expressions for lengths and $\alpha$-values of periodic
orbits. For example, for the case $n=4$ and $m=1$ one obtains
\begin{equation} \label{e20}
\mbox{sn}(\frac{K}{2}) = \frac{1}{\sqrt{1+\sqrt{1-\kappa^2}}}
\; , \hspace{6ex}
\mbox{Z}(\frac{K}{2}) = \frac{1}{2} 
\frac{\kappa^2}{1 + \sqrt{1 - \kappa^2}} \; .
\end{equation}
From this follows that the periodic orbits have the
conserved quantity $\alpha = b^4/(a^2+b^2)$ and length
$l = 4 \sqrt{a^2 + b^2}$.

\newpage

\noindent \underline{The case $\alpha < 0$:}

\vspace{2mm} \noindent
For $\alpha < 0$ the condition for periodic orbits is
\begin{equation} \label{e21}
\mbox{F}(\arcsin \sqrt{\frac{b^2}{b^2 - \alpha}},\frac{1}{\kappa}) =
\frac{2 m}{n} \, \mbox{F}(\frac{\pi}{2},\frac{1}{\kappa}) =
\frac{2 m}{n} \, \mbox{K}(\frac{1}{\kappa}) \; ,
\end{equation}
with the restriction that $n$ has to be an even integer, since
the bounces at the boundary occur alternately in the upper and
lower half of the ellipse. This requirement is a consequence
of the definition of $I_u$ in Eqs.\ (\ref{e8}) and (\ref{e9})
which is the action for only half a cycle if $\alpha <0$.

The tori of periodic orbits can be labelled by $n=4,6,\dots$, and
$1 \le m < n/2$, where the integers $n$ and
$m$ are the number of reflections of an orbit and its libration
number, respectively.
Eq.\,(\ref{e21}) has a solution for all these values of $n$ and $m$
in the range $0 < \alpha < -\infty$. However, if the value
of $\alpha$ for a solution is smaller than $-c^2$, then the
torus is complex, as is the modulus $1/\kappa$. If one decreases
the ratio $b/a$ then the torus becomes real. This happens when
\begin{equation} \label{e22}
\frac{b}{a} = \sin \frac{m \pi}{n} \; ,
\end{equation}
which follows from Eq.\,(\ref{e21}) with $\alpha = -c^2$ 
and F($\varphi$,0) = $\varphi$.
For $\alpha=-c^2$ the torus has zero extension
and coincides with the orbit along the minor axis. Thus
all tori arise through bifurcations of the stable orbit 
(and its repetitions) when the ratio $b/a$ is decreased
from its starting value $b/a=1$ in a circle. 

We discuss again the case $m=n/2$ which corresponds to the
periodic orbit along the major axis and its repetitions.
Also for $\alpha <0$ there exists no solution for 
Eq.\,(\ref{e21}) if $m=n/2$ as can be seen by 
using (\ref{e15}). This yields
\begin{equation} \label{e23}
\lim_{\alpha \rightarrow 0} \left[
\mbox{F}(\frac{\pi}{2},\frac{1}{\kappa}) -
\mbox{F}(\arcsin \sqrt{\frac{b^2}{b^2 - \alpha}},\frac{1}{\kappa}) 
\right] = \mbox{F}(\arctan \frac{c}{b},1) 
= \log \frac{a+b}{c} \; .
\end{equation}

The inverted form of condition (\ref{e21}) is given by
\begin{equation} \label{e24}
\sqrt{\frac{b^2}{b^2 - \alpha}} 
= \mbox{sn}(\frac{2m}{n} \mbox{K}) \; ,
\end{equation}
where the modulus of the functions sn and K is now $1/\kappa$.

The lengths of the periodic orbits are given by
\begin{eqnarray} \label{e25}
l_{n,m} &=& \frac{2 \pi}{\sqrt{E}} (n I_u + m I_v) \nonumber \\
&=& \frac{2 n a b}{\sqrt{b^2 - \alpha}} 
- 2 n c \,
\mbox{E}(\arcsin \sqrt{\frac{b^2}{b^2 - \alpha}},\frac{1}{\kappa})
+ 4 m c                \, \mbox{E}(\frac{\pi}{2},\frac{1}{\kappa}) 
\nonumber \\
&=& \frac{2 n a b}{\sqrt{b^2 - \alpha}} - 2 n c \,
\mbox{Z}(\frac{2m}{n} \mbox{K}) \; ,
\end{eqnarray}
where Eqs.\,(\ref{e10}) and (\ref{e21}) have been used.

We give again explicit expressions for the case $n=4$ and
$m=1$. The values of sn(K/2) and Z(K/2) are now
given by (\ref{e20}) with $\kappa$ replaced by $1/\kappa$.
It follows that the periodic orbits have the
conserved quantity $\alpha = -b^4/(a^2-b^2)$ and length
$l = 4 a^2 / \sqrt{a^2 - b^2}$.

Detailed illustrations of properties of periodic orbits in
the ellipse can be found in \cite{ONO90}.

\subsection{Quantum mechanics of the elliptical billiard}
\label{secquant}

The Schr\"odinger equation in elliptical coordinates has
the following form (in dimensionless units $\hbar = 2 m = 1$)

\begin{equation} \label{e26}
- \left( \frac{\partial^2}{\partial u^2} + 
         \frac{\partial^2}{\partial v^2} \right)
\psi(u,v) = E c^2 (\cosh^2 u - \cos^2 v) \psi(u,v) \; .
\end{equation}
Writing $\psi(u,v) = \psi_1(v) \psi_2(u)$ this equation
separates into two ordinary differential equations
\begin{eqnarray} \label{e27}
 \psi_1''(v) + (b - h^2 \cos^2  v) \psi_1(v)  &=& 0 
\\ \label{e28}
-\psi_2''(u) + (b - h^2 \cosh^2 u) \psi_2(u)  &=& 0 \; ,
\end{eqnarray}
where
$h = \sqrt{E} c$ and $b = E (c^2 + \alpha)$.
We use here and in the following the notation of
Morse and Feshbach \cite{MF53}. The first equation (\ref{e27})
is Mathieu's equation. For every value of $h$ there is
a countable number of values of b for which
it has periodic solutions of period $\pi$ or
$2 \pi$. These values are denoted by $\mbox{be}_r(h), r=0,1,\dots$,
and $\mbox{bo}_r(h), r=1,2,\dots$. The corresponding solutions
are the Mathieu functions $\mbox{Se}_r(h,z)$ and 
$\mbox{So}_r(h,z)$, where $z=\cos v$, $0<v<\pi$. They are
even and odd, respectively, with respect to reflection
on the x-axis ($v \rightarrow -v$).

The second equation (\ref{e28}) is Mathieu's equation for 
an imaginary argument:
With the values $\mbox{be}_r$ and $\mbox{bo}_r$ for the constant $b$, 
the solutions of this equation that are regular at $u=0$ are the
radial Mathieu functions of the first kind 
$\mbox{Je}_r(h,z)$ and $\mbox{Jo}_r(h,z)$ where $z = \cosh u$. The
energy eigenvalues of the Schr\"odinger equation follow from
the condition
\begin{eqnarray} \label{e29}
\mbox{Je}_r(h,\cosh U) &=& 0 \; , \; \; r=0,1,\dots   \nonumber \\
\mbox{Jo}_r(h,\cosh U) &=& 0 \; , \; \; r=1,2,\dots
\end{eqnarray}
where $U = \mbox{arctanh}(b/a)$. The functions Je and Jo have 
the properties $[d \mbox{Je}_r(h,\cosh u) / du]_{u=0} = 0$ and
$\mbox{Jo}_r(h,\cosh u)|_{u=0} = 0$.

The solutions can be separated into the four symmetry classes of
the elliptical billiard. These symmetry classes will be denoted by
two letters, where the first denotes the boundary condition on
the x-axis and the second that on the y-axis. For example,
$DN$ denotes the symmetry class of wave functions which satisfies 
Dirichlet boundary conditions on the x-axis and Neumann boundary
condition on the y-axis.

The solutions of the Schr\"odinger equation for all four symmetry
classes are, up to a normalization constant,
\begin{equation} \label{e30}
\begin{array}{ll}
DD: \; \; \; \mbox{So}_{2r+2}(h,\cos v) 
\, \mbox{Jo}_{2r+2}(h,\cosh u) \; \; \; \; \; &
DN: \; \; \; \mbox{So}_{2r+1}(h,\cos v) 
\, \mbox{Jo}_{2r+1}(h,\cosh u) \\
ND: \; \; \; \mbox{Se}_{2r+1}(h,\cos v)
\, \mbox{Je}_{2r+1}(h,\cosh u) \; \; \; \; \; &
NN: \; \; \; \mbox{Se}_{2r}(h,\cos v)
\, \mbox{Je}_{2r}(h,\cosh u) \; , 
\end{array}
\end{equation}
where $r=0,1,\dots$. In the limit $h \rightarrow 0$
these solutions reduce to the solutions of the circular
billiard, i.\,e.\ to a product of trigonometric and
Bessel functions \cite{MF53}.

A numerical examination of the energy spectrum of the
elliptical billiard in dependence on the ratio of the two
half-axis can be found in \cite{TFB89}.

\subsection{The semiclassical approximation for the 
            elliptical billiard}

In this section we derive a semiclassical trace formula for the
spectral density of the elliptical billiard. We follow the method
of Berry and Tabor \cite{BT76} and start with the EBK-quantization.
The semiclassical spectral density which is obtained from it
is then reexpressed by applying the Poisson summation formula
to it. From this representation the trace formula for the ellipse
is derived.

\subsubsection{The EBK-quantization}

The EBK-quantization conditions for the elliptical billiard 
have been examined in \cite{KR60}. They are given by
\begin{equation} \label{sc1}
\begin{array}{lllll}
\alpha > 0: \; \; \; \; & I_u = n_u + \frac{3}{4} \; , & 
\; \; n_u = 0,1,\dots \; \; \; \; & I_v = |n_v| \; , &
\; \; n_v \in \mbox{\bf Z} \\
\alpha < 0: \; \; \; \; & 2 I_u = n_u + 1 \; , &
\; \; n_u = 0,1,\dots \; \; \; \; & I_v = n_v + \frac{1}{2} \; , &
\; \; n_v = 0,1,\dots 
\end{array}
\end{equation}
The condition for $I_v$ and $\alpha >0$ corresponds to periodic
boundary conditions, and the factor of $2$ which appears
in the condition for $I_u$ and $\alpha < 0$ is due to the fact
that $I_u$ is the action for only half a cycle. The formulation
of the EBK-conditions for the full ellipse is not very convenient.
If the $\alpha$-value of a semiclassical state changes sign
(as a consequence of changing the ratio $b/a$),
then the state is described by different quantum numbers 
than before. For that reason, it is advantageous to formulate
the semiclassical quantization for a half-ellipse where this
problem does not appear. The semiclassical levels of the full 
ellipse are then obtained by adding the spectra of two half-%
ellipses which have Dirichlet ($D$) or Neumann ($N$) boundary 
conditions on the $x$-axis, respectively, (and Dirichlet boundary
conditions on the remaining arc). For these two systems the 
EBK-conditions have the form 
\begin{equation} \label{sc2}
\begin{array}{lllll}
D: \; \; \; \; & \alpha > 0: \; \; \; \; & 
I_u = n_u + \frac{3}{4}      \; \; \; \; & 
I_v = n_v + 1                \; \; \; \; &
n_u, n_v = 0,1,\dots         \; \; \; \; \\[4pt]
               & \alpha < 0: \; \; \; \; & 
I_u = n_u + 1                \; \; \; \; & 
I_v = n_v + \frac{1}{2}      \; \; \; \; &
n_u, n_v = 0,1,\dots         \; \; \; \; \\[4pt]
N: \; \; \; \; & \alpha > 0: \; \; \; \; & 
I_u = n_u + \frac{3}{4}      \; \; \; \; & 
I_v = n_v                    \; \; \; \; &
n_u, n_v = 0,1,\dots         \; \; \; \; \\[4pt]
               & \alpha < 0: \; \; \; \; & 
I_u = n_u + \frac{1}{2}      \; \; \; \; & 
I_v = n_v + \frac{1}{2}      \; \; \; \; &
n_u, n_v = 0,1,\dots         \; \; \; \; \\
\end{array}
\end{equation}
As one can see, the quantization conditions are discontinuous
at $\alpha = 0$, and as a consequence the semiclassical
quantization is inaccurate near $\alpha = 0$. This was examined
in detail in \cite{AA87a,AA87b}. It was found that near
$\alpha =0$ there are sometimes two different semiclassical
approximations for one quantum state, and sometimes there is
no semiclassical approximation. A remedy to this problem is
the application
of a uniform approximation near $\alpha = 0$, which yields
quantization conditions in terms of a noninteger Maslov index
that interpolates smoothly between the cases
$\alpha < 0$ and $\alpha > 0$. This uniform approximation
has been derived in \cite{AA87a} by expanding the potential
terms in Eqs.\ (\ref{e27}) and (\ref{e28}) up to second
order in $u$ and $v$. The solutions of the corresponding
differential equations are parabolic cylinder functions.
By matching the asymptotic form of these solutions with
the EBK-solutions one obtains a uniform approximation for
the Maslov phase. As a consequence the conditions (\ref{sc2})
are replaced by
\begin{equation} \label{sc3}
\begin{array}{llll}
D: \; \; & \ds
I_u = n_u + \frac{1}{2} + 
\frac{\beta_A(\bar{\alpha})}{\pi} \; \; & \ds
I_v = n_v + \frac{2 \beta_A(-\bar{\alpha})}{\pi} \; \; &
n_u, n_v = 1, 2, \dots            \\[6pt]
N: \; \; & \ds
I_u = n_u + \frac{1}{2} + 
\frac{\beta_S(\bar{\alpha})}{\pi} \; \; & \ds
I_v = n_v + \frac{2 \beta_S(-\bar{\alpha})}{\pi} \; \; &
n_u, n_v = 1, 2, \dots \; ,
\end{array}
\end{equation}
where $\bar{\alpha} = \alpha \sqrt{E} /(2 c)$ and
\begin{equation} \label{sc4}
\beta_S(\bar{\alpha}) = \theta - \frac{\pi}{4} + 
\frac{\bar{\alpha}}{2} \log|\bar{\alpha}| - 
\frac{\bar{\alpha}}{2} - \frac{1}{2}
\arg \Gamma(\frac{1}{2} + i \bar{\alpha}) \; ,
\end{equation}
\begin{equation} \label{sc5}
\theta = \arctan[\sqrt{1 + e^{2 \pi \bar{\alpha}}} 
+ e^{\pi \bar{\alpha}}] \; ,
\end{equation}
and $\beta_A(\bar{\alpha})$ is given by (\ref{sc4}) with 
the replacement of $\theta$ by $\pi - \theta$. The original
conditions (\ref{sc2}) are recovered with the limiting
values of $\beta_S$ and $\beta_A$
\begin{equation} \label{sc6}
\bar{\alpha} \rightarrow +\infty: \; \; \beta_S, \beta_A 
\rightarrow \frac{\pi}{4} \; \; \; \; \; \; \; \; \; \; \; \;
\bar{\alpha} \rightarrow -\infty: \; \; \beta_S \rightarrow 0
\; , \; \; \beta_A \rightarrow \frac{\pi}{2} \; .
\end{equation}
With this uniform approximation for the Maslov phase the
previous ambiguity of the semiclassical quantization is
removed. A further advantage of the uniform approximation
is that it removes a semiclassical degeneracies of energy
levels with $\alpha > 0$. According to the quantization
condition (\ref{sc2}) the even and odd states ($D$ and $N$)
have the same energy values (except for the lowest even
state). This degeneracy is a semiclassical degeneracy 
since the true quantum levels are not degenerate.

An alternative possibility for treating the semiclassical
influence of the separatrix consists in the inclusion
of complex tunneling orbits. This is done for the
ellipse in \cite{DWW96}.

\subsubsection{The spectral density}

We derive now a trace formula for the spectral density of
the elliptical billiard in terms of the periodic orbits of
the system. We follow the method of Berry and Tabor
\cite{BT76} and apply the Poisson summation formula to the
semiclassical spectral density which is obtained from the
EBK-quantization. We do this here for the two cases of a
half-ellipse with Dirichlet or Neumann boundary conditions
on the $x$-axis. The spectral density is given by
\begin{eqnarray} \label{sc8}
d(E) &=& \sum_{n_u, n_v = 0}^\infty 
\delta (E - E_{n_u,n_v})
\\ \nonumber
&=& \sum_{n, m = -\infty}^\infty 
\int_0^\infty \! \mbox{d} I_u \, \int_0^\infty \! 
\mbox{d} I_v \,
\exp\{2 \pi i n (I_u - \frac{\nu_u}{4}) 
+ 2 \pi i m (I_v - \frac{\nu_v}{4}) \}
\, \delta (E - E_{n_u,n_v}) \; ,
\end{eqnarray}
where after the application of the Poisson summation formula
the integration variables have been changed from $n_u$ and 
$n_v$ to $I_u = n_u - \nu_u/4$ and $I_v = n_v - \nu_v/4$.
The values $E_{n_u,n_v}$ denote the energies which are
determined by the EBK-quantization conditions (\ref{sc3}) for
either $D$- or $N$-boundary conditions. For simplicity
of notation we do not write an additional index for the
boundary conditions. The quantities $\nu_u$ and $\nu_v$
are the Maslov indices. They are approximated by the
uniform approximation given in (\ref{sc3}). After a
change of variables from $I_u$ and $I_v$
to $\alpha$ and $E$ one obtains 
\begin{equation} \label{sc9}
d(E) = \sum_{n, m = -\infty}^\infty 
\int_{-c^2}^{b^2} \! \mbox{d} \alpha \, J(\alpha,E)
e^{2 \pi i n (I_u - \nu_u/4) + 2 \pi i m (I_v - \nu_v/4)} \; ,
\end{equation}
where $J(\alpha,E)$ is the Jacobian of the transformation.
For values of $\alpha$ and $E$ where 
$\bar{\alpha} = \alpha \sqrt{E} /(2 c)$ is not small one
can neglect the $\alpha$-dependence of the Maslov phases
and the Jacobian is given by 
\begin{equation} \label{sc10}
J(\alpha,E)
  = \left| \frac{\partial I_u}{\partial E} 
           \frac{\partial I_v}{\partial \alpha} - 
           \frac{\partial I_v}{\partial E}
           \frac{\partial I_u}{\partial \alpha} \right|
  = \frac{1}{2 E} 
    \left| I_u \frac{\partial I_v}{\partial \alpha} - 
           I_v \frac{\partial I_u}{\partial \alpha} \right| \; ,
\end{equation}
which is a function of $\alpha$ only.
\begin{figure}[th] 
\begin{center}
\mbox{\epsfig{file=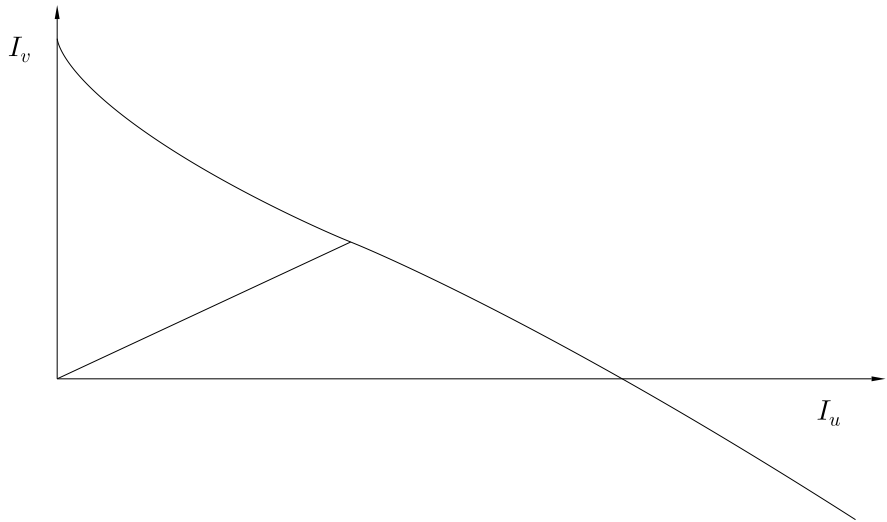,width=12cm,
bbllx=20pt, bblly=0pt, bburx=650pt, bbury=380pt}}
\end{center}
\caption{The curve $(I_v(\protect\alpha),I_u(\protect\alpha))$
for a half-ellipse with axis-ratio 
$\protect\frac{b}{a}=\protect\frac{9}{11}$. The action $I_v$
is negative if $\protect\alpha < -c^2$. The line from the
origin marks the point corresponding to $\protect\alpha = 0$.}
\end{figure}

We discuss in the following the different contributions
to the integrals in (\ref{sc9}). In doing this we
assume that the eccentricity of the ellipse is not
very small. If it is very small then the unstable
and stable orbits along the two axis of the ellipse
cannot be treated as isolated orbits, since they
result from the break-up of a torus of the circular
billiard. An investigation of the semiclassical
contributions of these orbits for small eccentricity
was carried out in \cite{Cre96}.

\subsubsection{The area contribution}

The term with $n=0$ and $m=0$ in (\ref{sc9}) and (\ref{sc8})
is the only one with a non-oscillatory integrand. It has
been shown in general that this term corresponds to the
area term of Weyl's law \cite{BT76}
\begin{equation} \label{sc11}
d_{0,0}(E) = 
\int_0^\infty \! \mbox{d} I_u \, \int_0^\infty \! 
\mbox{d} I_v \, \delta (E - E_{n_u,n_v})
= \frac{A}{4 \pi} \; ,
\end{equation}
where $A$ is the area of the billiard system. For a
half-ellipse $A=\pi a b / 2$.

\subsubsection{Contributions of orbits with $\alpha > 0$}
\label{secposa}

For all other terms the integrand in (\ref{sc9}) is
oscillatory and the main contribution comes from values
of $\alpha$ where the phase of the exponential function
is stationary. We consider first contributions from
$\alpha >0$ and approximate them by a stationary phase
approximation. We neglect a possible $\alpha$-%
dependence of the Maslov phase, i.\,e.\ we assume 
that the energy is large enough so that
$\bar{\alpha} = \alpha \sqrt{E} /(2 c)$ is not
small at a stationary point and the Maslov index can
be considered constant. Then the Jacobian is
given by (\ref{sc10}) and we drop its second argument
since it is energy-independent.
The integrals in (\ref{sc9}) can then be considered
to be integrals over the energy surface $E=\mbox{const.}$
of the billiard system. This energy surface is plotted
in figure~1. The stationary points are those
points in this figure where the slope of the curve is
rational. The curvature of the energy surface is positive
for $\alpha > 0$ and negative for $\alpha < 0$.

The stationary phase condition is
\begin{equation} \label{sc12}
\frac{\partial}{\partial \alpha} (n I_u + m I_v) = 0 \; ,
\end{equation}
and it coincides with the condition (\ref{e12})
for periodic orbits. It has a solution for $n=3,4,\dots$ and
$1 \le m < n/2$ and for the corresponding negative values of
$m$ and $n$. The term for negative values $(-n,-m)$ is the
complex conjugate of the term for $(n,m)$ as can be seen 
from Eq.\,(\ref{sc9}) since $J$ is real. For that reason
we calculate the contributions for positive values of 
$n$ and $m$ and add at the end a complex conjugate contribution.
The result of the stationary phase approximation for a term
corresponding to a pair $(n,m)$ is 
\begin{equation} \label{sc13}
J(\alpha_{n,m}^e) \frac{
\exp \{2 \pi i n (I_u - \frac{\nu_u^e}{4}) 
     + 2 \pi i m (I_v - \frac{\nu_v^e}{4}) + i \frac{\pi}{4} \} }{
\sqrt{n \frac{\partial^2 I_u}{\partial \alpha^2} + 
      m \frac{\partial^2 I_v}{\partial \alpha^2}} } \; ,
\end{equation}
where $\alpha_{n,m}^e$ is the value of $\alpha$ at the stationary
point. The superscript $e$ denotes that we consider in this section
contributions of periodic orbits with elliptical caustic.

The second derivatives of the actions are given by
\begin{eqnarray} \label{sc14}
\frac{\partial^2 I_u}{\partial \alpha^2} 
&=& -\frac{\kappa^2}{4 \alpha c^2} \left(
I_u - \frac{\sqrt{E} a b}{\pi \sqrt{b^2 - \alpha}} \right) 
\nonumber \\
\frac{\partial^2 I_v}{\partial \alpha^2} 
&=& -\frac{\kappa^2}{4 \alpha c^2} I_v \; .
\end{eqnarray}
This relation is also valid for $\alpha <0$. At the stationary
points one further has
\begin{equation} \label{sc15}
J(\alpha_{n,m}^e) = \frac{l_{n,m}^e}{4 \pi n \sqrt{E}} 
\frac{\partial I_v}{\partial \alpha} \; ,
\end{equation}
which follows from Eqs.\,(\ref{e12}) and (\ref{e18}).

Combining all results one obtains the joint contribution
of the pairs $(n,m)$ and $(-n,-m)$
\begin{equation} \label{sc16}
d_{n,m}^e(E) = \sqrt{\frac{2 \alpha_{n,m}^e}{\pi k}}
\frac{l_{n,m}^e \mbox{F}\,(\frac{\pi}{2},\kappa_{n,m}^e)}{
      n \pi \sqrt{\frac{2 n a b}{\sqrt{b^2 -\alpha_{n,m}^e}}
- l_{n,m}^e}}
\cos ( k l_{n,m}^e - \frac{\pi}{2} n \nu_u^e
- \frac{\pi}{2} m \nu_v^e  + \frac{\pi}{4} ) \; ,
\end{equation}
where $k=\sqrt{E}$ is the wave number.
The results for the two half-ellipses are identical since the
Maslov indices are $\nu_u^e=3$ and $\nu_v^e=4$ for $D$,
and $\nu_u^e=3$ and $\nu_v^e=0$ for $N$ and thus the argument
of the cosine differs by a multiple of $2 \pi$.

For the derivation of Eq.\ (\ref{sc16}) we assumed that 
the parameter $\bar{\alpha}$ is large at a stationary
point. If this is not the case then the approximation
is more complicated. Then one has to take into
account the $\alpha$-dependence of the uniform 
approximation for the Maslov-phase in the exponent when
one determines the stationary points. This leads to a
replacement of condition (\ref{sc12}) by
\begin{equation} \label{sc12b}
\frac{\partial}{\partial \alpha} (n n_u + m n_v) = 0 \; ,
\end{equation}
where $n_u$ and $n_v$ are expressed as functions of
$\alpha$ by Eqs. (\ref{sc3}). The same modification
applies also to tori with small negative values of
$\bar{\alpha}$. Condition (\ref{sc12b}) does
not correspond to the condition for periodic
orbits (\ref{sc12}) any more, but in the limit
$E \rightarrow \infty$ the previous stationary phase
condition (\ref{sc12}) is recovered. This is similar 
to the treatment of whispering gallery orbits 
in the circular billiard \cite{Uss94}. 
(For contributions of whispering gallery orbits in the
stadium billiard see \cite{Tan96}.)
Their semiclassical description involves effective 
lengths which are different from the lengths of
periodic orbits.
We do not further discuss the
modifications due to Eq.\ (\ref{sc12b}), a detailed
investigation of contributions of tori near the
separatrix is carried out in \cite{DWW96}. Finally,
we remark that we did not consider corrections for
whispering gallery orbits with $\alpha \approx b^2$
in this section. These corrections go beyond the
EBK-quantization \cite{Uss94}.

\subsubsection{Contributions of orbits with $\alpha < 0$}
\label{secnega}

In the case of negative $\alpha$ the main contributions
come again from the stationary points of the integrand
in (\ref{sc9}), and as before we assume the $\bar{\alpha}$
is not small at these points. 
However, now the stationary points can also
lie outside the integration range (when the corresponding
torus is complex) or very close to the lower end of the
integration range. For these cases a stationary phase
approximation is not appropriate and one has to use a 
uniform approximation instead. We apply here the uniform
approximation of Berry and Tabor \cite{BT76} for integrals
with a stationary point that can lie near the boundary or 
outside the integration range. This approximation can
be written in the form \cite{Ric82}
\begin{eqnarray} \label{sc17}
\int_{\alpha_0}^\infty \! \mbox{d} \alpha \, g(\alpha) \,
e^{i k f(\alpha)} &=& \frac{g(\alpha^*) 
\sqrt{2 \pi i \beta}}{\sqrt{k\,|f''(\alpha^*)|}}
e^{i k f(\alpha^*)} \left[ \Theta(\alpha^*-\alpha_0)
+ \sqrt{\frac{i \beta}{2 \pi}} \mbox{sign}(\alpha^*-\alpha_0)
\int_{\Lambda}^\infty \! \mbox{d} X \frac{1}{X^2}
e^{\frac{i \beta}{2} X^2} \right]
\nonumber \\ &&
+ \frac{i}{k} \frac{g(\alpha_0)}{f'(\alpha_0)} 
e^{i k f(\alpha_0)} \; ,
\end{eqnarray}
where  
$\beta = \mbox{sign}(f''(\alpha^*))$, 
$\Lambda = \sqrt{2 k \beta (f(\alpha_0) - f(\alpha^*))}$,
$\alpha^*$ is determined by $f'(\alpha^*) = 0$,
and $\Theta(\alpha)$ denotes the Heaviside theta function. 
The terms on the right-hand side of Eq.\,(\ref{sc17})
can be interpreted in the following way: the term multiplying the 
$\Theta$-function is the stationary phase approximation of the
integral, the last term is the contribution from the boundary
(which can be obtained by an integration by parts), and the
remaining term is an interference term between both. 
As will be seen in the following section the sum over all
boundary contributions gives the Gutzwiller expression
for the semiclassical contribution of the stable orbit along
the minor axis of the billiard and its repetitions. When the
ratio $b/a$ of the two half-axis is decreased then new tori
of periodic orbits arise through bifurcations of the
stable orbit (and its repetitions).
At the bifurcation point the new tori coincide with the stable
orbit and have zero extension, and as $b/a$ is decreased further
they separate from the stable orbit. This is expressed by
formula (\ref{sc17}). Far away from the bifurcation 
the interference term can be neglected and the torus is
well separated from the stable orbit. However, near the
bifurcation, the contributions of the torus and
the stable orbit cannot be separated, they rather 
give a joint contribution. At the bifurcation the
Gutzwiller contribution of one repetition of the stable
orbit diverges and this divergence is cancelled by a
similar divergence of the interference term.

We consider in this section only the stationary phase term
and the interference term in Eq.\,(\ref{sc17}). The sum over
the boundary contributions is performed in the next section.  
The stationary phase condition is again given by 
Eq.\,(\ref{sc12}) which has a solution for $n=3,4,\dots$ and
$1 \le m < n/2$ and for the corresponding negative values of
$m$ and $n$ (since we consider a half-ellipse $n$ does not
have to be even). As before we calculate only the contributions
for positive values of $n$ and $m$ and add at the end a
complex conjugate contribution.

The second derivatives of the actions are given 
by Eq.\,(\ref{sc14}) and the value of the Jacobian
at a stationary point by
Eq.\,(\ref{sc15}) with the difference that now the
quantities are given a superscript $h$ to signify the
contributions of tori with a hyperbolical caustic. The value
of $\beta$ in (\ref{sc17}) is $-1$ since the curvature of
the energy surface is negative for negative values of
$\alpha$. Inserting these values into (\ref{sc17}) one
obtains
\begin{eqnarray} \label{sc18}
d_{n,m}^h(E) &=& \sqrt{\frac{2 |\alpha_{n,m}^h|}{\pi k}}
\frac{l_{n,m}^h \mbox{F}\,(\frac{\pi}{2},
\frac{1}{\kappa_{n,m}^h})}{
      n \pi \sqrt{(\frac{2 n a b}{\sqrt{b^2 -\alpha_{n,m}^h}}
      - l_{n,m}^h) (\kappa_{n,m}^h)^2 }}
\nonumber \\ &&
\left[ \Theta(\alpha_{n,m}^h + c^2) \,
\cos ( k l_{n,m}^h - \frac{\pi}{2} n \nu_u^h
- \frac{\pi}{2} m \nu_v^h  - \frac{\pi}{4} )
\right.         
\nonumber \\ &&    
\left. + \frac{\mbox{sign}(\alpha_{n,m}^h + c^2)}{\sqrt{2 \pi}} 
\int_\Lambda^\infty \! \mbox{d} X \,
\frac{\sin(k l_{n,m}^h - X^2/2 
- \frac{\pi}{2} n \nu_u^h - \frac{\pi}{2} m \nu_v^h 
)}{X^2}   \right] \; ,
\end{eqnarray}
where $\Lambda = \sqrt{2 k (l_{n,m}^h - 2nb)}$ and
the Maslov indices are $\nu_u^h=4$ and $\nu_v^h=2$ for $D$,
and $\nu_u^h=2$ and $\nu_v^h=2$ for $N$.

\subsubsection{The contribution of the stable orbit}

We sum now over all contributions from the boundary 
$\alpha = - c^2$ of the integrals in (\ref{sc9}). These 
boundary contributions are given by the last term in
(\ref{sc17}) and they exist for all values of $n$ and $m$,
even when there is no stationary point for these
values. The only exception is when
both values are equal to zero because then the integrand
is non-oscillatory and yields no boundary contribution.
It has been shown in general by Richens that the
contribution of a stable periodic orbit can be obtained
by summing over all boundary contributions 
in an integrable system \cite{Ric82}.

The values of the actions and their derivatives at
$\alpha = -c^2$ are given by
\begin{equation} \label{sc19}
\begin{array}{ll} \ds
I_u|_{\alpha=-c^2} = \frac{b \sqrt{E}}{\pi} \; \; \; \; & \ds
\left. \frac{\partial I_u}{\partial \alpha} 
\right|_{\alpha=-c^2} = -\frac{\sqrt{E}}{2 \pi c}
\arcsin \frac{b}{a} \; \; \; \; \\[4pt] \ds
I_v|_{\alpha=-c^2} = 0 \; \; \; \; & \ds
\left. \frac{\partial I_v}{\partial \alpha} 
\right|_{\alpha=-c^2} =  \frac{\sqrt{E}}{2 c} \, ,
\end{array}
\end{equation}
and $J(-c^2) = b/(4 \pi c)$.
Summing over all boundary contributions one obtains
\begin{eqnarray} \label{sc20}
d_s(E) &=& \sum_{n,m=-\infty}^\infty \!\!\!\!\!\!{}' \, \, 
\frac{i b}{4 \pi k} \frac{\exp\{ 2 i n k b - \frac{i \pi}{2}
n \nu_u^h - \frac{i \pi}{2} m \nu_v^h \}}{
\pi m - n \arcsin \frac{b}{a}}
\nonumber \\ &=&
\sum_{n=1}^\infty \frac{b}{\pi k} 
\frac{\sin(2 n k b - \frac{\pi}{2} n \nu_u^h)}{
2 \sin(n \arcsin \frac{b}{a} )} \; ,
\end{eqnarray}
where the prime at the first sum indicates that the term
$(n,m)=(0,0)$ is omitted. Furthermore, the relation
\begin{equation} \label{sc21}
\sum_{m=-\infty}^\infty \frac{(-1)^m}{z - \pi m}
= \frac{1}{\sin(z)}
\end{equation}
has been used \cite{AS65} and the fact that $\nu_v^h=2$ for $D$ 
and $N$.

The semiclassical contribution of the stable orbit and its
repetitions $d_s(E)$ is
divergent when $n \arcsin (b/a) = m \pi$. This coincides
with the condition Eq.\,(\ref{e22}) for the appearance of a
new torus.

\subsubsection{The perimeter contribution}

At the boundary of the integration $\alpha = b^2$ one
obtains also a contribution to the energy density. It is
obtained by the approximation
\begin{equation} \label{sc22}
\mbox{boundary contribution of} \left\{
\int_{-\infty}^{\alpha_1} \! \mbox{d} \alpha \, g(\alpha) \,
e^{i k f(\alpha)} \right\} \approx 
- \frac{i}{k} \frac{g(\alpha_1)}{f'(\alpha_1)} 
e^{i k f(\alpha_1)} \; ,
\end{equation}
which follows from an integration by parts. Near $\alpha = b^2$
the actions are expanded in powers of $\varepsilon = b^2 - \alpha$:
\begin{eqnarray} \label{sc23}
I_u &\approx& \frac{\sqrt{E} \sqrt{\varepsilon^3}}{3 \pi a b}  
\nonumber \\
I_v &\approx& \frac{2 a \sqrt{E}}{\pi} \mbox{E}\,(\frac{\pi}{2},
\frac{c}{a}) - \varepsilon \frac{\sqrt{E}}{\pi a}
\mbox{F}\,(\frac{\pi}{2},\frac{c}{a}) \; .
\end{eqnarray}
It follows that
$J(\alpha) \approx - I_v (2 E)^{-1} \partial I_u / \partial \alpha$
and it's leading term is proportional to $\varepsilon^{1/2}$
for small $\varepsilon$. For terms with $m \neq 0$ 
the exponent in the integral (\ref{sc9}) depends in leading order
linearly on $\varepsilon$ and the corresponding boundary
contribution vanishes. The contributions of the terms with $m = 0$
and $n \neq 0$ are given by
\begin{eqnarray} \label{sc24}
d_p(E) &=& - \sum_{n=-\infty}^\infty \!\!\!{}' 
\int^{b^2} \! \mbox{d} \alpha \,
\left(\frac{I_v}{2 E} 
\frac{\partial I_u}{\partial \alpha} \right)
\exp \{ 2 \pi i n I_u - \frac{i \pi}{2} n \nu_u^e \}
\nonumber \\
&=& \left. \mbox{\bf Re} \sum_{n=1}^\infty
\frac{i I_v}{2 \pi n E} \exp \{ 2 \pi i n I_u - 
\frac{i\pi}{2} n \nu_u^e \} \right|_{\alpha=b^2}
\nonumber \\
&=& \frac{I_v}{2 \pi E} \sum_{n=1}^\infty 
\frac{\sin( \frac{\pi}{2} n \nu_u^e)}{n}
\nonumber \\
&=& - \frac{a \mbox{E}\,(c/a)}{4 \pi k} \; ,
\end{eqnarray}
where $\nu_u^h = 3$ has been inserted for $D$ and $N$. The
term (\ref{sc24}) is the perimeter contribution $-L/(8 \pi k)$ of
the outer arc of the half-ellipse to the mean spectral density,
where $L= 2 a \mbox{E}(e)$.

\subsubsection{The contribution of the unstable orbit}

We now consider the question how the contribution of the 
unstable periodic orbit can be obtained. This orbit has
an $\alpha$-value of zero, and at this point
the original EBK-quantization is discontinuous and 
inaccurate. It has been first pointed out by Bogomolny
that the semiclassical contribution of the unstable
orbit can be obtained by using the uniform approximation 
for the semiclassical phase near the separatrix
\cite{Bog88}. We follow his method in this section.

In the context of the uniform approximation it is more
convenient to work with the $\bar{\alpha}$-variable 
instead of the $\alpha$-variable. For that reason
we start again with equation (\ref{sc8}) and change
the integration variable from $I_u$ and $I_v$ to
$\bar{\alpha}$ and $E$ and obtain
\begin{equation} \label{sc25}
d^{N,D}(E) = \sum_{n, m = -\infty}^\infty 
\int \! \mbox{d} \bar{\alpha} \, \bar{J}(\bar{\alpha},E)
\exp \{2 \pi i n (I_u - \frac{1}{2} -
       \frac{  \beta_{S,A}( \bar{\alpha})}{\pi}) 
     + 2 \pi i m (I_v - 
       \frac{2 \beta_{S,A}(-\bar{\alpha})}{\pi}) \} \; ,
\end{equation}
where the new Jacobian is given by
\begin{eqnarray} \label{sc26}
\bar{J}(\bar{\alpha},E) 
&=& \left| \frac{\partial n_u}{\partial E}
           \frac{\partial n_v}{\partial \bar{\alpha}} -
           \frac{\partial n_v}{\partial E}
           \frac{\partial n_u}{\partial \bar{\alpha}} \right|
\\ \nonumber
&=& \left| 
      \frac{\partial I_u}{\partial E}
\left(\frac{\partial I_v}{\partial \bar{\alpha}} + \frac{2}{\pi}
      \frac{\partial \beta_{S,A} (-\bar{\alpha})}{
            \partial \bar{\alpha}} \right) -
      \frac{\partial I_v}{\partial E}
\left(\frac{\partial I_u}{\partial \bar{\alpha}} - \frac{1}{\pi}
      \frac{\partial \beta_{S,A} (\bar{\alpha})}{
            \partial \bar{\alpha}} \right)
\right| \; .
\end{eqnarray}
We evaluate now the contributions to the integrals in (\ref{sc25})
from the vicinity of $\bar{\alpha}=0$, and we restrict
to contributions with $n=2m$ since these terms are the ones
which correspond to the unstable orbit as has been discussed
in section \ref{secclass}.

The derivative of $\beta_{S,A}(\bar{\alpha})$ with respect to
$\bar{\alpha}$ is given by
\begin{equation} \label{sc27}
\frac{\partial}{\partial \bar{\alpha}} \beta_{S,A}(\bar{\alpha})
= \pm \frac{\pi}{4 \cosh(\pi \bar{\alpha})} 
+ \frac{1}{2} \log|\bar{\alpha}| 
- \frac{1}{4} \Psi(\frac{1}{2} + i \bar{\alpha})
- \frac{1}{4} \Psi(\frac{1}{2} - i \bar{\alpha}) \; , 
\end{equation}
where $\Psi(z)$ is the logarithmic derivative of the 
$\Gamma$-function \cite{AS65}. 
For small values of $\bar{\alpha}$ one further has 
\begin{eqnarray} \label{sc28}
I_u &\approx& \frac{\sqrt{E}(a-c)}{\pi} - \frac{\bar{\alpha}}{2 \pi}
- \frac{\bar{\alpha}}{2 \pi}
 \log \frac{8 c \sqrt{E}}{|\bar{\alpha}|}
+ \frac{\bar{\alpha}}{\pi} \log \frac{a+c}{b} 
\nonumber \\
I_v &\approx& \frac{2 \sqrt{E} c}{\pi} + \frac{\bar{\alpha}}{\pi}
+ \frac{\bar{\alpha}}{\pi} \log \frac{8 c \sqrt{E}}{|\bar{\alpha}|}
\; .
\end{eqnarray}
With these equations the Jacobian $\bar{J}$ is evaluated near
$\bar{\alpha}=0$
\begin{equation} \label{sc29}
\bar{J}(\bar{\alpha},E) 
\approx \frac{1}{2 \pi \sqrt{E}} \left( \frac{a}{\pi}
\log(8 c \sqrt{E}) - \frac{2 c}{\pi} \log \frac{a+c}{b} 
\pm \frac{a}{2 \cosh(\pi \bar{\alpha})} 
- \frac{a}{2 \pi} [ \Psi(\frac{1}{2} + i \bar{\alpha}) 
+ \Psi(\frac{1}{2} - i \bar{\alpha}) ] \right) \; .
\end{equation}
In expression (\ref{sc29}) for $\bar{J}$ we have set all
appearing $\bar{\alpha}$-values equal to zero except in
the argument of the cosh- and $\Psi$-function. The reason
for this will become clear in the following. Inserting 
Eqs.\ (\ref{sc28}) and (\ref{sc29}) into (\ref{sc25})
one obtains 

\begin{eqnarray} \label{sc30}
d_u^{N,D}(E) &=& 2 \mbox{\bf Re} \sum_{m=1}^\infty
\int \! \mbox{d} \bar{\alpha} \,
\frac{(-1)^m}{2 \pi \sqrt{E}} \left( 
\frac{a}{\pi} \log(8 c \sqrt{E}) 
- \frac{2 c}{\pi} \log \frac{a+c}{b}
\pm \frac{a}{2 \cosh(\pi \bar{\alpha})} \right.
\nonumber \\ && \left.
- \frac{a}{2 \pi} [
\Psi(\frac{1}{2} + i \bar{\alpha}) +
\Psi(\frac{1}{2} - i \bar{\alpha}) ] \right) 
\exp \left\{4 i m a \sqrt{E} + 4 i m \bar{\alpha} \log
\frac{a+c}{b} \right\} \; ,
\end{eqnarray}
where the relations 
$\beta_S(\bar{\alpha})+\beta_S(-\bar{\alpha}) = \pi/4$ and
$\beta_A(\bar{\alpha})+\beta_A(-\bar{\alpha}) = 3 \pi/4$ have
been used in the exponent. 

As one can see, the exponent in (\ref{sc30}) depends linearly
on $\bar{\alpha}$ and for that reason one would not expect
to obtain a contribution of the order of an
unstable orbit from $\bar{\alpha}=0$. The reason that
one does obtain this contribution nevertheless is that the
Jacobian in (\ref{sc29}) has poles with 
$\mbox{\bf Re}\, \bar{\alpha} = 0$. More precisely,
$\Psi(1/2 + i \bar{\alpha})$ has poles at $i(l+1/2)$ 
with residuum $i$ for $l=0,1,\dots$, 
$\Psi(1/2 - i \bar{\alpha})$ has poles at $-i(l+1/2)$ 
with residuum $-i$ for $l=0,1,\dots$, 
and $1/\cosh(\pi \bar{\alpha})$ has poles at $i(l+1/2)$
with residuum $-i (-1)^l/\pi$ for $l \in \mbox{Z$\!\!$Z}$.
The contribution of the unstable orbit is obtained
by closing the integration contour in the upper half
plane and using Cauchy's theorem
\begin{eqnarray} \label{sc31}
d_u^{N,D}(E) &=& 2 \mbox{\bf Re} \sum_{m=1}^\infty
\left\{ 
\sum_{l=0}^\infty \frac{a (-1)^m}{2 \pi k}
\exp\{ 4 i m a k - 2 m \log \frac{a+c}{b} 
- 4 m l \log \frac{a+c}{b} \} 
\right. \nonumber \\ &&  \left. 
\pm \sum_{l=0}^\infty \frac{a (-1)^m (-1)^l}{2 \pi k}
\exp\{ 4 i m a k - 2 m \log \frac{a+c}{b} 
- 4 m l \log \frac{a+c}{b} \} 
\right\}
\nonumber \\
&=& 2 \mbox{\bf Re} \sum_{m=1}^\infty
\left\{ 
\frac{a (-1)^m}{2 \pi k} 
\frac{\exp\{4 i m a k \}}{(\frac{a+c}{b})^{2m}
- (\frac{a+c}{b})^{-2m}} 
\pm \frac{a (-1)^m}{2 \pi k} 
\frac{\exp\{4 i m a k \}}{(\frac{a+c}{b})^{2m}
+ (\frac{a+c}{b})^{-2m}}
\right\} 
\nonumber \\ 
&=& \sum_{m=1}^\infty \frac{a}{\pi k}
\cos(4 m a k - \pi m)
\left\{ \frac{1}{2 \sinh(2 m \, \mbox{arcsinh} \frac{c}{b})}
\pm     \frac{1}{2 \cosh(2 m \, \mbox{arcsinh} \frac{c}{b})} 
\right\} \; .
\end{eqnarray}
This is the contribution of the unstable orbit which runs
along the $x$-axis that is part of the boundary of
the half-ellipse.

\subsubsection{The joint contribution}
\label{secjoint}

Combining all results of the previous sections
and adding the contributions for $D$ and $N$ one 
obtains the trace formula for the full ellipse.
We give it here for the level density 
$\hat{d}(k) = 2 k d(E)$ in terms of the wave number $k$
\begin{eqnarray} \label{sc32}
\hat{d}(k) &=& \bar{d}(k) + \sum_{n=3}^\infty
\sum_{m=1}^{[\frac{n-1}{2}]}
\sqrt{\frac{2 k \alpha_{n,m}^e}{\pi}}
\frac{4 l_{n,m}^e \mbox{F}\,(\frac{\pi}{2},\kappa_{n,m}^e)}{
      n \pi \sqrt{\frac{2 n a b}{\sqrt{b^2 -\alpha_{n,m}^e}}
- l_{n,m}^e}}
\cos ( k l_{n,m}^e - \frac{3 \pi}{2} n + \frac{\pi}{4} ) 
\nonumber \\ && + 
\sum_{\stackrel{n=4}{\scriptscriptstyle n \, 
even}}^\infty \sum_{m=1}^{[\frac{n-1}{2}]}
\sqrt{\frac{2 k |\alpha_{n,m}^h|}{\pi}}
\frac{4 l_{n,m}^h \mbox{F}\,(\frac{\pi}{2},
\frac{1}{\kappa_{n,m}^h})}{
      n \pi \sqrt{(\frac{2 n a b}{\sqrt{b^2 -\alpha_{n,m}^h}}
      - l_{n,m}^h) (\kappa_{n,m}^h)^2 }}
%\nonumber \\ && \times
\left[ \Theta(\alpha_{n,m}^h + c^2) \,
\cos ( k l_{n,m}^h - \pi m - \frac{\pi}{4} )
\right.
\nonumber \\ &&
+ \left. \frac{\mbox{sign}(\alpha_{n,m}^h + c^2)}{\sqrt{2 \pi}} 
\int_\Lambda^\infty \! \mbox{d} X \,
\frac{\sin(k l_{n,m}^h - X^2/2 - \pi m )}{X^2}   \right]
\nonumber \\ && +
\sum_{m=1}^\infty \frac{4 b}{\pi}
\frac{\sin(4 m k b)}{2 \sin(2 m \, \arcsin \frac{b}{a} )} +
\sum_{m=1}^\infty \frac{4 a}{\pi}
\frac{\cos(4 m k a - m \pi)}{
2 \sinh(2 m \, \mbox{arcsinh} \frac{c}{b} )} \; ,
\end{eqnarray}
where $\bar{d}(k)$ is the mean spectral density 
and $\Lambda = \sqrt{2 k (l_{n,m}^h - 2 n b)}$.

The same method can be used to obtain also the trace
formula for quarter ellipses with the four different
boundary conditions given in section \ref{secquant}.
We give here only the result. The modified EBK-quantization
conditions are
\begin{equation} \label{sc33}
\begin{array}{llll}
DD: \; \; & \ds
I_u = n_u + \frac{1}{2} + 
\frac{\beta_A(\bar{\alpha})}{\pi} \; \; & \ds
\frac{I_v}{2} = n_v + \frac{1}{2}
+ \frac{\beta_A(-\bar{\alpha})}{\pi} \; \; &
n_u, n_v = 1, 2, \dots            \\[6pt]
DN: \; \; & \ds
I_u = n_u + \frac{1}{2} + 
\frac{\beta_A(\bar{\alpha})}{\pi} \; \; & \ds
\frac{I_v}{2} = n_v
+ \frac{\beta_A(-\bar{\alpha})}{\pi} \; \; &
n_u, n_v = 1, 2, \dots            \\[6pt]
ND: \; \; & \ds
I_u = n_u + \frac{1}{2} + 
\frac{\beta_S(\bar{\alpha})}{\pi} \; \; & \ds
\frac{I_v}{2} = n_v + \frac{1}{2} 
+ \frac{\beta_S(-\bar{\alpha})}{\pi} \; \; &
n_u, n_v = 1, 2, \dots            \\[6pt]
NN: \; \; & \ds
I_u = n_u + \frac{1}{2} + 
\frac{\beta_S(\bar{\alpha})}{\pi} \; \; & \ds
\frac{I_v}{2} = n_v + \frac{\beta_S(-\bar{\alpha})}{\pi} \; \; &
n_u, n_v = 1, 2, \dots \; .
\end{array}
\end{equation}
where $I_v/2$ with definition (\ref{e10}) is the value of the
action in a quarter ellipse. The trace formula for a quarter
ellipse is given by
\begin{eqnarray} \label{sc34}
\hat{d}(k) &=& \bar{d}(k) + \sum_{n=2}^\infty
\sum_{m=\frac{1}{2}}^{\frac{n-1}{2}}
\sqrt{\frac{2 k \alpha_{n,m}^e}{\pi}}
\frac{l_{n,m}^e \mbox{F}\,(\frac{\pi}{2},\kappa_{n,m}^e)}{
      n \pi \sqrt{\frac{2 n a b}{\sqrt{b^2 -\alpha_{n,m}^e}}
- l_{n,m}^e}}
\cos ( k l_{n,m}^e - \frac{\pi}{2} n \nu_u^e
- \pi m \nu_v^e + \frac{\pi}{4} ) 
\nonumber \\ && + 
\sum_{n=2}^\infty \sum_{m=\frac{1}{2}}^{\frac{n-1}{2}}
\sqrt{\frac{2 k |\alpha_{n,m}^h|}{\pi}}
\frac{l_{n,m}^h \mbox{F}\,(\frac{\pi}{2},
\frac{1}{\kappa_{n,m}^h})}{
      n \pi \sqrt{(\frac{2 n a b}{\sqrt{b^2 -\alpha_{n,m}^h}}
      - l_{n,m}^h) (\kappa_{n,m}^h)^2 }}
\nonumber \\ && \times 
\left[ \Theta(\alpha_{n,m}^h + c^2) \,
\cos ( k l_{n,m}^h - \frac{\pi}{2} n \nu_u^h
- \pi m \nu_v^h - \frac{\pi}{4} )
\right.         
\nonumber \\ && +
\left. \frac{\mbox{sign}(\alpha_{n,m}^h + c^2)}{\sqrt{2 \pi}} 
\int_\Lambda^\infty \! \mbox{d} X \,
\frac{\sin(k l_{n,m}^h - X^2/2 - \frac{\pi}{2} n \nu_u^h
- \pi m \nu_v^h)}{X^2}   \right]
\nonumber \\ && +
\sum_{m=1}^\infty \frac{b}{\pi} \left\{
\frac{\sin(2 m k b - \frac{\pi}{2} m \nu_u^h)}{
      2 \sin(m \, \arcsin \frac{b}{a} )} +
\frac{\cos(2 m k b - \frac{\pi}{2} m \nu_u^h 
      - \frac{\pi}{2}(\nu_v^h - 1))}{
      2 \cos(m \, \arcsin \frac{b}{a} )} \right\}
\nonumber \\ && 
+ \sum_{m=1}^\infty \frac{a}{\pi} \left\{
\frac{\cos(2 m k a - \frac{\pi}{2} m (\nu_u + \nu_v))}{
2 \sinh(m \, \mbox{arcsinh} \frac{c}{b} )} +
\frac{\cos(2 m k a - \frac{\pi}{2} m(\nu_u + \nu_v) 
      - \frac{\pi}{2}(\nu_u^h - 2))}{
2 \cosh(m \, \mbox{arcsinh} \frac{c}{b} )} \right\} \; ,
\end{eqnarray}
where $m$ runs now over half-integers in the sums over tori,
$\Lambda = \sqrt{2 k (l_{n,m}^h - 2 n b)}$,
and the Maslov indices are given by
\begin{equation} \label{sc35}
\begin{array}{lllll}
  DD: \; \; &
\nu_u^e = 3 \; \; &
\nu_v^e = 4 \; \; &
\nu_u^h = 4 \; \; &
\nu_v^h = 3
\\ DN: \; \; &
\nu_u^e = 3 \; \; &
\nu_v^e = 2 \; \; &
\nu_u^h = 4 \; \; &
\nu_v^h = 1
\\ ND: \; \; &
\nu_u^e = 3 \; \; &
\nu_v^e = 2 \; \; &
\nu_u^h = 2 \; \; &
\nu_v^h = 3
\\ NN: \; \; &
\nu_u^e = 3 \; \; &
\nu_v^e = 0 \; \; &
\nu_u^h = 2 \; \; &
\nu_v^h = 1 \; .
\end{array}
\end{equation}
The sum $(\nu_u + \nu_v)$ in the last line of Eq.\ (\ref{sc34})
can be taken either at $\alpha > 0$ or $\alpha < 0$ since it
is invariant.

We remark that in Eqs.\ (\ref{sc32}) and (\ref{sc34}) we did not
include modifications for tori close to the separatrix 
that were discussed in section \ref{secposa}.

\subsection{Numerical results}

\begin{figure}[thbp] 
\begin{center}
\mbox{\epsfig{file=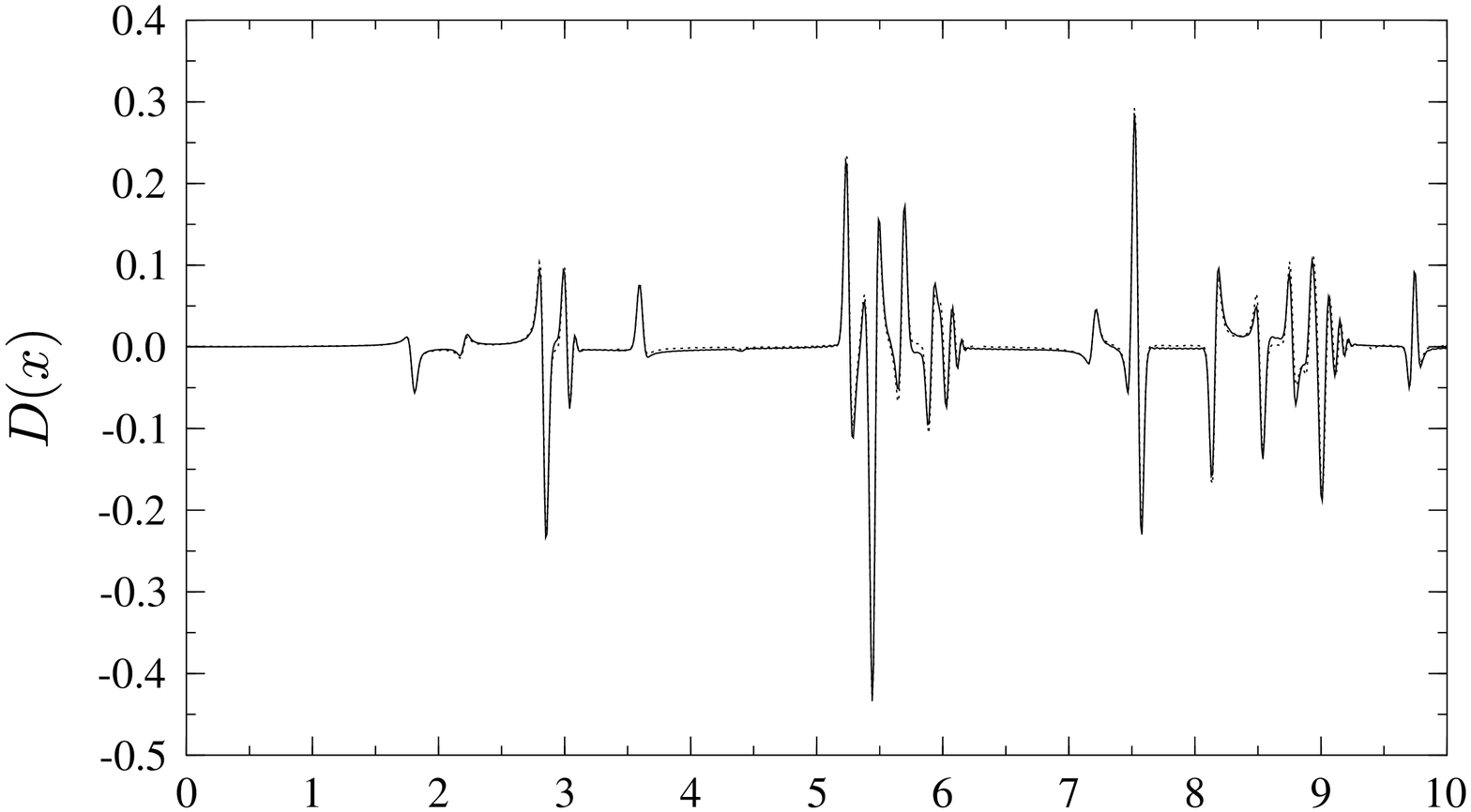,width=16cm,
bbllx=0pt, bblly=-21pt, bburx=750pt, bbury=380pt}}
\end{center}
\caption{The Fourier cosine transforms $D_{\scriptscriptstyle QM}(x)$
(full) and $D_{\scriptscriptstyle SC}$ (dashed) of the level density,
calculated from quantum energies and periodic orbits, respectively.}
\vspace*{-0.2cm}
\begin{center}
\mbox{\epsfig{file=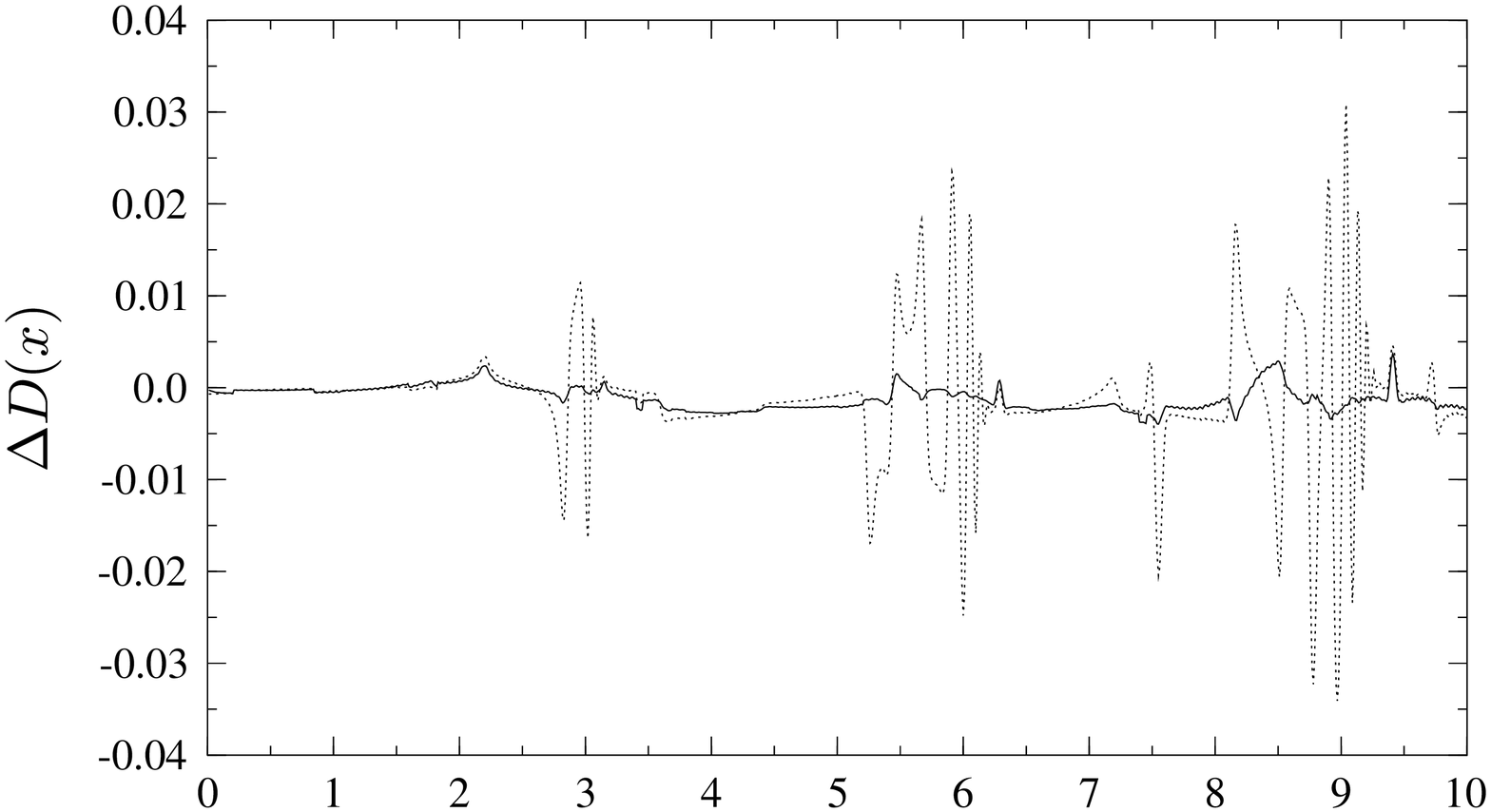,width=16cm,
bbllx=0pt, bblly=-21pt, bburx=750pt, bbury=380pt}}
\end{center}
\caption{The differences $D_{\scriptscriptstyle QM}(x) - 
D_{\scriptscriptstyle SC}(x)$ and $D_{\scriptscriptstyle EBK}(x) - 
D_{\scriptscriptstyle SC}(x)$ between Fourier cosine transforms
of the level density.}
\end{figure}
\begin{figure}[thbp] 
\begin{center}
\mbox{\epsfig{file=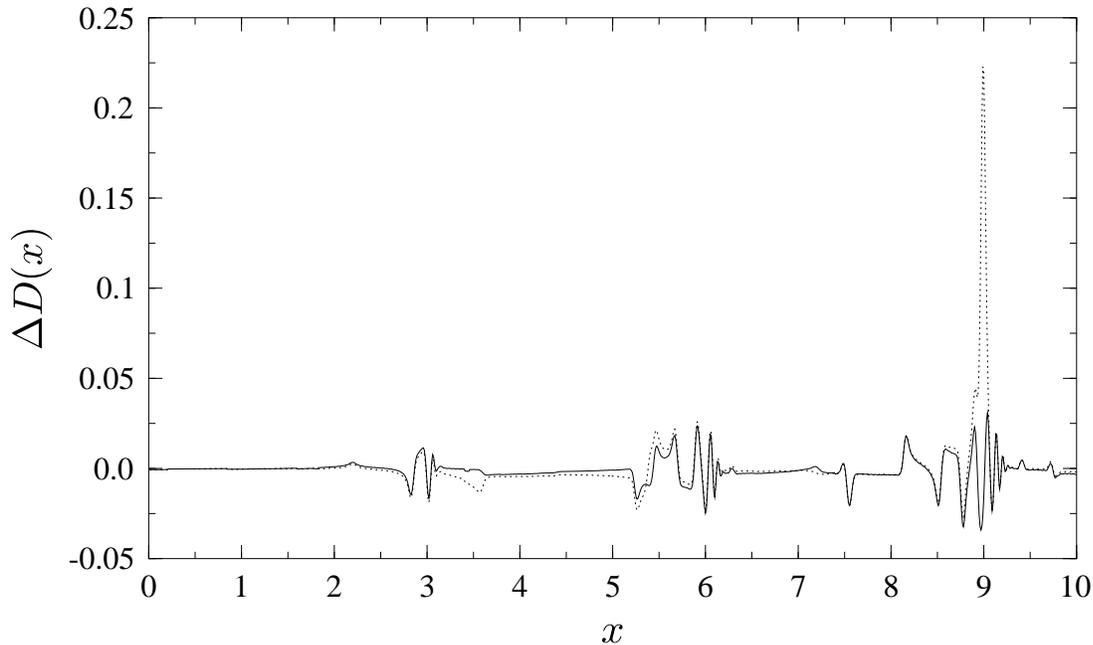,width=16cm,
bbllx=0pt, bblly=-21pt, bburx=750pt, bbury=380pt}}
\end{center}
\caption{The differences $D_{\scriptscriptstyle QM}(x) - 
D_{\scriptscriptstyle SC}(x)$ between Fourier cosine transforms
of the level density with (full) and without (dashed) contributions
from interference terms to $D_{\scriptscriptstyle SC}(x)$.}
\end{figure}
In this section we compare the semiclassical approximation
of the last section with exact quantum
results. For this purpose we consider a Fourier cosine
transform of the oscillatory part of the level density
with a Gaussian damping term and a cut-off
\begin{equation} \label{num1}
D(x) = \frac{1}{k_{max}} \int_0^{k_{max}} \! \mbox{d}k \,
[ d(k) - \bar{d}(k) ] \, \cos(k x) \, 
\exp(-z \frac{k^2}{k_{max}^2}) \; .
\end{equation}
This function has peaks in the vicinity of the lengths of
periodic orbits.

The calculations were carried out for a quarter ellipse
with $a=1.1$ and $b=0.9$ and with Dirichlet boundary 
conditions on all sides. The damping factor was chosen
to be $z=5$. The function $D(x)$ was 
evaluated in three different ways and the results are
denoted $D_{\scriptscriptstyle QM}(x)$,
$D_{\scriptscriptstyle SC}(x)$
and $D_{\scriptscriptstyle EBK}(x)$. The first function
$D_{\scriptscriptstyle QM}(x)$ was obtained from the quantum
energies. These energy levels were determined up to energy
$E_{max}=20\,000$ by solving the two Mathieu equations 
(\ref{e27}) and (\ref{e28}) numerically, which corresponds to
$1\,198$ energy levels. The second function 
$D_{\scriptscriptstyle SC}(x)$ was determined by using
Eq.\,(\ref{sc34}) with all periodic orbits up to length
$l=12$ and $n=30$, and the third function 
$D_{\scriptscriptstyle EBK}(x)$ was determined from the
semiclassical energy levels that are solutions of 
the EBK-conditions (\ref{sc33}).

In figure~2 the results for the quantum spectrum
and the trace formula, $D_{\scriptscriptstyle QM}(x)$
and $D_{\scriptscriptstyle SC}(x)$, are compared. Both
curves are in good agreement and can hardly be distinguished.
For that reason we plot the difference between
both curves in figure~3 (dashed line).
One can see that the semiclassical error is approximately
one order of magnitude smaller than the function $D(x)$. 
Figure~3  shows also the difference between 
$D_{\scriptscriptstyle EBK}(x)$ and 
$D_{\scriptscriptstyle SC}(x)$ (full line) which is 
much smaller than the difference between
$D_{\scriptscriptstyle QM}(x)$ and 
$D_{\scriptscriptstyle SC}(x)$. This shows that the
error which was introduced by deriving the trace formula 
from the EBK-energies with stationary phase and uniform
approximations is much smaller than the original error
of the EBK-quantization.
This cannot necessarily be expected because the EBK-quantization
condition, the trace formula and the stationary
phase approximations which connect both approximations 
are all only valid in leading order of $\hbar$.
A similar result was observed previously for a circular
billiard with a singular magnetic flux line \cite{RBMBM96}.
The smallness of $D_{\scriptscriptstyle EBK}(x) - 
D_{\scriptscriptstyle SC}(x)$ shows also that the
modifications for contributions of tori near the
separatrix due to Eq.\ (\ref{sc12b}) are not large
in the range where the numerical examination was
carried out.

We also show that it is important to include
complex orbits, i.\,e.\ the interference terms, in 
(\ref{sc34}). In figure~4 the difference 
$D_{\scriptscriptstyle QM}(x) - D_{\scriptscriptstyle SC}(x)$
is plotted where $D_{\scriptscriptstyle SC}(x)$ was calculated
once with (full line) and once without (dashed line) the
interference terms. One can see that the semiclassical 
error is much bigger in the second case, especially near
$x=9$. This is due to a torus of complex orbits with $n=5$
and $m=1.5$ which is close to becoming real. It becomes
real for $b/a \approx 0.809$ which is near the present value of
$b/a = 11/9 \approx 0.818$.

\section{The oval billiard}

We discuss now how the semiclassical 
approximation for the elliptical billiard has to be modified
when the ellipse experiences a small perturbation
which makes the system non-integrable. We consider
in particular a perturbation which consists in a deformation
of the ellipse into an oval-shaped billiard system
introduced by Berry \cite{Ber81}. 
It has a parameterization which expresses the radius of
curvature $R$ of the boundary as a function of the angle
$\Psi$ between the tangent vector and the $x$-direction
\begin{equation} \label{o1}
R(\Psi) = 1 + \delta \cos(2 \Psi) \; ,
\end{equation}
where $\Psi$ lies in the range $[0,2 \pi)$. From this definition
the dependence of the $x$- and $y$-coordinates on
$\Psi$ follows as \cite{Ber81}
\begin{eqnarray} \label{o2}
x(\Psi) &=& \sin(\Psi) + \frac{\delta}{2} \sin(\Psi) + 
\frac{\delta}{6} \sin(3 \Psi)
\nonumber \\ 
y(\Psi) &=& -\cos(\Psi) + \frac{\delta}{2} \cos(\Psi) - 
\frac{\delta}{6} \cos(3 \Psi) \, .
\end{eqnarray}
The oval has an area $A = \pi - \pi \delta^2 / 6$,
a perimeter $L = 2 \pi$, two half-axes of lengths
$a = 1 + \delta/3$ and $b = 1 - \delta/3$, and it
is a deformation of an ellipse of order $\delta^2$. 
The billiard can as well be considered
as a perturbation of the circle. Then the 
deformation is of order $\delta$.

\begin{figure}[p] 
\begin{center}
\mbox{\epsfig{file=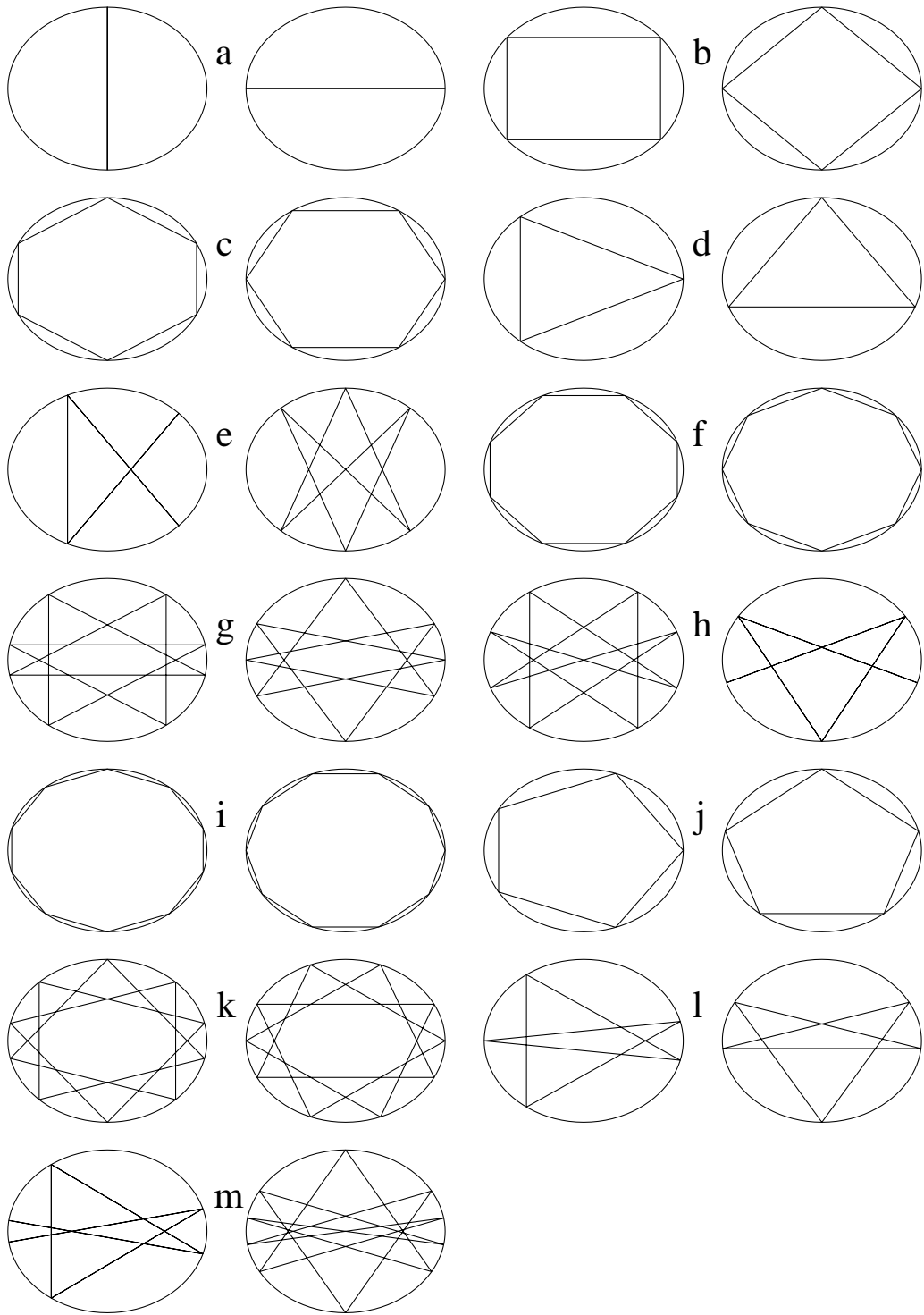,width=14.5cm,
bbllx=-50pt, bblly=-5pt, bburx=530pt, bbury=785pt}}
\end{center}
\caption{Periodic orbits in the oval billiard
for $\delta = 0.3$. The desymmetrized versions
of these orbits in the quarter oval represent
all orbits which are reflected up to five times
on the oval-shaped boundary.}
\end{figure}
Numerical examinations were performed for the deformation
parameter $\delta = 0.3$ for which the half-axes of the
oval billiard have the same lengths as those of the
elliptical billiard in the last section.
We consider again a desymmetrized version of the
billiard which consists of a quarter oval with
Dirichlet boundary conditions on all sides. 
For this system the shortest periodic orbits were determined.
Figure~5 represents all periodic orbits which
are reflected up to five times on the oval-shaped
boundary of the quarter oval. We choose to
present them by corresponding orbits in the
full oval since their structure can be seen more clearly 
there. All these orbits appear in pairs which
are labelled by letters. With the exception of the
pair $a$ all pairs of orbits
result from the break-up of tori of the ellipse.
Since the perturbation is small, no subsequent
bifurcations of orbits occur in this length regime.

The differences in the lengths of orbits within a pair 
are shown in Table 1. With the exception
of the first pair the length differences are very small.
For that reason, the orbits cannot be treated as isolated
orbits in a semiclassical approximation if the energy
is not rather high. We discuss in the following the
different uniform approximations by which the
contributions of the pairs to the semiclassical
trace formula can be treated. Most of the pairs
($b,c,d,f,g,i,j,k,l$) result from a break-up of
tori of the elliptical billiard with $\alpha > 0$, 
i.\,e.\ from tori which have a confocal ellipse
as caustic. Their contribution can be treated
by the uniform approximation for the generic
break-up of tori, that will be described in the
next section. 
\begin{table}[htb] 
\begin{center}
\begin{tabular}{|c|c|c|r|c|}
\hline 
pair & $\bar{l}$ & $\Delta l$ & $\Delta t$  \phantom{xl} 
\rule{0ex}{2.7ex} & type \\[0.3ex]
\hline
a  & 2.00000 & 0.20000 & \rule{0ex}{2.7ex} $0.52747$ & \\
b  & 2.83548 & 0.00705 &  $0.03875$ & e \\
c  & 3.00259 & 0.00051 &  $0.00265$ & e \\
d  & 5.23473 & 0.00249 &  $0.01208$ & e \\
e  & 5.43636 & 0.00025 & $-0.00035$ & h \\
f  & 3.06282 & 0.00005 & $ 0.00017$ & e \\
g  & 7.49347 & 0.00169 &  $0.02427$ & e \\
h  & 7.54064 & 0.00112 & $-0.01176$ & h \\
i  & 3.09101 & 0.00001 & $ 0.00001$ & e \\
j  & 5.88609 & 0.00007 & $ 0.00031$ & e \\
k  & 8.12883 & 0.00029 &  $0.00288$ & e \\
l  & 9.70973 & 0.00149 &  $0.07798$ & e \\
m  & 9.72132 & 0.00135 & $-0.06491$ & h \\[0.3ex]
\hline
2b & 5.67096 & 0.01411 & \rule{0ex}{2.7ex} $0.61722$ & e \\
3b & 8.50644 & 0.02116 &  $3.17563$ & e \\[0.3ex]
\hline
\end{tabular}
\end{center}
\caption{Properties of pairs of orbits in the quarter oval. 
The different columns show the mean length $\bar{l}$ the
length difference $\Delta l$ and the quantity 
$\Delta t = \protect\mbox{Tr\,} M_1 + 
\protect\mbox{Tr\,} M_2 - 4$ for a pair of orbits. The type-%
column specifies whether the pair arose from the break-up of a
torus of the ellipse with a confocal ellipse or a confocal
hyperbola as caustic, respectively.}
\end{table}

Other pairs ($e,h,m$) result
from the break-up of tori of the ellipse
with $\alpha < 0$, i.\,e.\ from tori with a
confocal hyperbola as caustic. These tori
all arose from a bifurcation of the stable
orbit in the ellipse, as the eccentricity
of the ellipse was increased starting from
the circle. These tori are not necessarily
well separated from the stable orbit in the
ellipse, as is expressed by the interference
term in (\ref{sc18}), and thus the usual formula
for the break-up of tori cannot be used.

The correct contributions of these pairs
to the trace formula can be obtained by
considering the oval not as a deformation
of the ellipse but as a deformation of
the circle. Then the pairs ($e,h,m$)
are obtained not from the break-up of
tori but from bifurcations of the stable
orbit along the vertical axis of the
billiard (the first orbit of the pair
$a$ in figure~5). This stable
orbit will be denoted by $a_s$ in the
following. In particular, the pair $e$ arises
from a bifurcation of the 3-fold repetition 
of the orbit $a_s$ at $\delta_{bif} = 3/13 \approx 0.2308$,
the pair $h$ from a bifurcation of the
4-fold repetition of $a_s$ at
$\delta_{bif} = (39-24 \sqrt{2})/41 \approx 0.1234$,
and the pair $m$ from a bifurcation of the
5-fold repetition of $a_s$ at
$\delta_{bif} = (63-24\sqrt{5})/121 \approx 0.0771$.
For the pair $e$ the bifurcation is shown in
figure~6 by plotting the orbits
for values of $\delta$ near  
$\delta_{bif} \approx 0.2308$. This bifurcation
results in four new orbits in the full oval, the right
orbit in figure~6 which can be traversed in
both directions and thus represents two orbits,
and the left orbit and its mirror image (obtained
by reflection on the $y$-axis) which is
not plotted.
\begin{figure}[thbp]
\begin{center}
\mbox{\epsfig{file=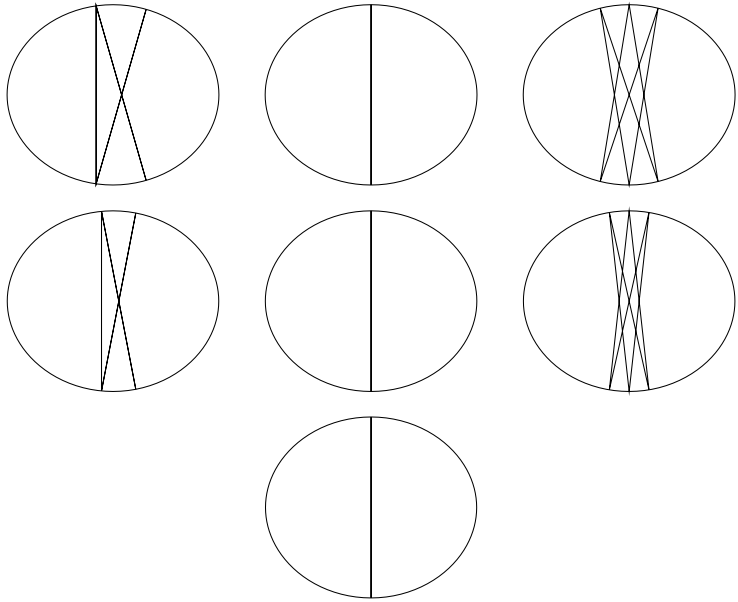,width=10cm,
bbllx=45pt, bblly=49pt, bburx=458pt, bbury=385pt}}
\end{center}
\caption{The period-tripling bifurcation which yields
the orbits $e$ of figure~5. The orbits
are shown for $\delta=0.230$, $\delta=0.235$ and
$\delta=0.240$ (from below), and the bifurcation occurs
at $\delta_{bif} = 3/13 \approx 0.231$.}
\end{figure}

It follows from the above considerations that
the contributions of the pairs $e$, $h$ and
$m$ are described by formulas
for bifurcations that are discussed in
section \ref{obif}. This is a very
general property. If one considers
an integrable system which depends
on an external parameter and in which
bifurcations occur that result in the
appearance of new tori, then in a slightly
perturbed system there are bifurcations
which result in a finite number of new
orbits. Thus the break-up of tori which
are close to a bifurcation is described
by formulas for bifurcations in which
a finite number of orbits is involved.

\subsection{Broken torus contribution}

If an integrable system is perturbed then all
tori of periodic orbits break up and only a
finite number of periodic orbits remain 
(except for degenerate cases). According to
the Poincar\'e-Birkhoff theorem the number
of remaining orbits is even, half them
are stable and half of them
are unstable. The Poincar\'e-Birkhoff theorem
does not specify the total number of
remaining orbits, but generally this number
is two \cite{LL92}. For the semiclassical
treatment of this case, i.\,e.\ the break-up
of a torus into one stable and one unstable
orbit, a uniform approximation was derived by 
Tomsovic, Grinberg and Ullmo \cite{TGU95,UGT96}.
This approximation
extends previous results of Ozorio de Almeida
\cite{Ozo86,Ozo88}. The formula can also be applied
to cases in which, due to the presence of
discrete symmetries, the break-up of a
torus results in $n$ stable and
$n$ unstable orbits with respectively
identical actions, periods, stabilities and
Maslov indices. 
The uniform approximation for the contributions
of these orbits to the level density can be written
in the following form
\begin{eqnarray} \label{bt1}
d_{bt}(E) &=& \frac{1}{2 \pi \hbar r} \,
\sqrt{\frac{2 \pi |\Delta S|}{\hbar}} \,
\mbox{\bf Re} \, \exp \left\{\frac{i}{\hbar} \bar{S} 
- \frac{i \pi}{2} \bar{\nu} \right\}
\nonumber \\ &&
\times \left\{
\left[ \frac{\bar{T} + \frac{\Delta T}{2}}{
\sqrt{|\mbox{Tr\,} M_u - 2|}}
+      \frac{\bar{T} - \frac{\Delta T}{2}}{
\sqrt{|\mbox{Tr\,} M_s - 2|}}
\right] \, \mbox{J}_0 \left( \frac{\Delta S}{\hbar} \right) 
\right. \nonumber \\ && 
+ \, i \,
\left[ \frac{T_u}{\sqrt{|\mbox{Tr\,} M_u - 2|}}
-      \frac{T_s}{\sqrt{|\mbox{Tr\,} M_s - 2|}}
\right] \, \mbox{J}_1 \left( \frac{\Delta S}{\hbar
} \right) 
\nonumber \\ && 
\left. - \phantom{\, i \,}
\left[ \frac{\frac{\Delta T}{2}}{\sqrt{|\mbox{Tr\,} M_u - 2|}}
-      \frac{\frac{\Delta T}{2}}{\sqrt{|\mbox{Tr\,} M_s - 2|}}
\right] \, \mbox{J}_2 \left( \frac{\Delta S}{\hbar} \right) 
\right\} \; .
\end{eqnarray}
In case of the break-up of a torus into $2n$ orbits due to the
presence of discrete symmetries the equation has to be multiplied
by $n$. The indices $u$ and $s$ correspond to the unstable and
stable orbit, respectively. Furthermore
\begin{equation} \label{bt2}
\bar{S}   = \frac{S_u + S_s}{2} \; , \; \; \; 
\Delta S  = \frac{S_u - S_s}{2} \; , \; \; \; 
\bar{T}   = \frac{T_u + T_s}{2} \; , \; \; \;
\Delta T  = \frac{T_u - T_s}{2} \; , \; \; \;
\bar{\nu} = \frac{\nu_u + \nu_s}{2} \; ,
\end{equation}
and $S$, $T$, $\nu$, $M$ and $r$ denote the action, period,
Maslov index, stability matrix and repetition number of an orbit,
respectively.

In the following we will use a slightly simplified version of
(\ref{bt1}) which is obtained by applying the relation
$\mbox{J}_2(z) = 2 \, \mbox{J}_1(z) / z - \mbox{J}_0(z)$. The 
term $ 2 \, \mbox{J}_1(z) / z$
yields a modification of the prefactor
of the $\mbox{J}_1$-Bessel function which is one order of
$\hbar$ smaller than the previous prefactor of $\mbox{J}_1$ 
in (\ref{bt1}). Since we consider only the leading order
semiclassical approximation we neglect this term and obtain 
\begin{equation} \label{bt3}
d_{bt}(E) = \frac{1}{\pi \hbar} \,
\sqrt{\frac{2 \pi |\Delta S|}{\hbar}} \, \left\{ 
\bar{A} \, \mbox{J}_0 \left( \frac{\Delta S}{\hbar} \right) 
\, \cos \left( \frac{\bar{S}}{\hbar} - \frac{\pi}{2} \bar{\nu} \right)
+ \Delta A \, \mbox{J}_1 \left( \frac{\Delta S}{\hbar} \right) 
\, \cos \left( \frac{\bar{S}}{\hbar} - \frac{\pi}{2} (\bar{\nu}-1) 
\right) \right\} \; ,
\end{equation}
where
\begin{equation} \label{bt4}
\bar{A} = \frac{1}{2 r}
\left[ \frac{T_u}{\sqrt{|\mbox{Tr\,} M_u - 2|}}
+      \frac{T_s}{\sqrt{|\mbox{Tr\,} M_s - 2|}} \right]
\; , \; \; \; 
\Delta A = \frac{1}{2 r}
\left[ \frac{T_u}{\sqrt{|\mbox{Tr\,} M_u - 2|}}
-      \frac{T_s}{\sqrt{|\mbox{Tr\,} M_s - 2|}} \right] \; .
\end{equation}
This approximation, as well as (\ref{bt1}), interpolates over
the whole regime from the torus contribution to the contributions
of isolated periodic orbits in the Gutzwiller approximation.
In the limit $\Delta S \rightarrow 0$ the mean amplitude
$\bar{A}$ diverges, but the product $\bar{A} \sqrt{|\Delta S|}$
remains finite, and the whole expression yields the
Berry-Tabor term for the semiclassical
contribution of a torus. In the opposite limit where
$\Delta S/\hbar >> 1$ one can replace the Bessel functions
by their leading order term for large arguments and
obtains the Gutzwiller approximation for the 
contributions of isolated orbits. Finally we note
that equation (\ref{bt3}) in combination with the
definitions (\ref{bt2}) and (\ref{bt4}) is invariant
under exchange of the labels $u$ and $s$. Thus the
definitions in (\ref{bt2}) and (\ref{bt4}) can also
be formulated in terms of an orbit 1 and an orbit 2
without specifying which of them is stable and which
unstable.

\subsection{Contribution of bifurcating orbits}
\label{obif}

There is only a finite number of generic forms in
which periodic orbits in two-dimensional systems
bifurcate. They are characterized by normal forms
that describe the classical motion in the vicinity
of a bifurcation. The different normal forms for generic
bifurcations of periodic orbits in two-dimensional
systems were derived by Meyer \cite{Mey70} and 
Bruno \cite{Brj70,Bru72} and are discussed in 
\cite{Ozo88,Arn93}.
Altogether there are six different kinds of
bifurcations which are classified according
to the lowest repetition number $m$ of a central orbit
for which the bifurcation occurs. 
(A special case is $m=1$ where
there is no orbit before the bifurcation.)
There is one kind of bifurcation
for $m=1$ up to $m=3$, respectively, two kinds
for $m=4$, and one for $m>4$. The corresponding
bifurcations are period-$m$-tupling bifurcations,
i.\,e.\ the primitive periods of the arising
periodic orbits are $m$ times the primitive period of 
the bifurcating central periodic orbit at the bifurcation.

We restrict the discussion in the following
to the case $m>4$. This kind of bifurcation is the
first that can be observed when an integrable
system is perturbed since the higher repetitions
of a stable orbit bifurcate earlier than
the lower repetitions. It has the following
form: a central stable periodic orbit bifurcates
and two new periodic orbits appear, one stable
and one unstable, while the central orbit remains
stable. The new appearing periodic orbits are
called satellite orbits. 

For the case $m>4$ a uniform approximation
was derived in \cite{Sie96} which interpolates over
the regime from the bifurcation
up to regions where the orbits can be
described in the Gutzwiller approximation.
This uniform approximation was obtained
by an extension of the method of Ozorio de
Almeida and Hannay \cite{OH87}.

The final formula in \cite{Sie96} was given
for bifurcating periodic orbits in
billiard systems and for bifurcating periodic orbits
without turning points in systems with potentials.
The derivation in \cite{Sie96} can also be
applied to orbits with turning points, since only
the Maslov index has to be changed.
The following uniform formula for the contributions
of bifurcating periodic orbits to the level
density includes all these cases

\begin{eqnarray} \label{bif1}
d_{bif}(E) &=& \frac{\Theta(\hat{\varepsilon})}{\pi \hbar} \,
\sqrt{\frac{2 \pi |\Delta S|}{\hbar}} \, \left\{ 
\bar{A} \, \mbox{J}_0 \left( \frac{\Delta S}{\hbar} \right) 
\, \cos \left( \frac{\bar{S}}{\hbar} - \frac{\pi}{2} \bar{\nu} \right)
+ \Delta A \, \mbox{J}_1 \left( \frac{\Delta S}{\hbar} \right) 
\, \cos \left( \frac{\bar{S}}{\hbar} - \frac{\pi}{2} (\bar{\nu}-1) 
\right) \right\}
\nonumber \\[0.3ex] && 
+ \mbox{sign}(\hat{\varepsilon}) \, \frac{\bar{A}}{\pi \hbar} \,
\sqrt{\frac{|\Delta S|}{\hbar}} \,
\int_{\Lambda}^\infty \! \mbox{d} X \, 
\frac{
\cos \left( \frac{\bar{S}}{\hbar} + \frac{\beta}{2} X^2 
-\frac{\pi}{2} (\nu - \beta) \right)}{X^2}
+ \frac{T_0}{\pi \hbar m r} 
\frac{
\cos \left( \frac{S_0}{\hbar} - \frac{\pi}{2} \nu_0 \right)
}{\sqrt{|\mbox{Tr\,} M_0 - 2|}} \; .
\end{eqnarray}
Here $\Lambda = \sqrt{2 \beta [S_0 - \bar{S}]/\hbar}$ and
$\hat{\varepsilon}$ is a parameter that is positive when all
three orbits are real, and negative when only the central
orbit is real. The values of $\nu$ and $\beta$
can be obtained from the Maslov indices of the (real) periodic
orbits, since 
$\nu_0 = \nu + \mbox{sign}(\hat{\varepsilon}) \cdot \beta$,
$\nu_u = \nu$ and $\nu_s = \nu - \beta$. The other
quantities are defined in (\ref{bt2}) and (\ref{bt4}).
The repetition number of the central stable orbit is
$mr$ where $m$ is the lowest repetition number for
which the bifurcation occurs. (When the $m$-th repetition
of a periodic orbit bifurcates, then the bifurcation
occurs also for all repetition numbers which are a
multiple of $m$.) Eq.\,(\ref{bif1}) is valid as long
as no subsequent bifurcations of the participating
periodic orbits occur before they can be considered
isolated. 
Note that formula (\ref{bif1}) in combination with 
definitions (\ref{bt2}) and (\ref{bt4})
is again invariant under exchange of 
the indices $u$ and $s$. The different terms in Eq.\ 
(\ref{bif1}) can be interpreted in the following way.
The first term with the two Bessel functions is a joint 
contribution of the two satellite orbits. It has a
form which is identical to the broken torus contribution
(\ref{bt3}). The last term in (\ref{bif1}) is the
semiclassical contribution of the central stable orbit,
and the remaining term is an interference term between
satellite orbits and central orbit.

The properties of the three orbits that participate
in the bifurcation are summarized in Table 2.
\begin{table}[htb] 
\begin{center}
\begin{tabular}{|c|lll|}
\hline
& $\hat{\varepsilon} < 0:$ & $\xi_0$ stable,   & $\nu_0 = \nu-1$ 
\rule{0ex}{2.7ex} \\[1.0ex]
& $\hat{\varepsilon} > 0:$ & $\xi_0$ stable,   & $\nu_0 = \nu+1$ \\
{\large $\beta = 1$}
&                          & $\xi_u$ unstable, & $\nu_u = \nu$   \\
&                          & $\xi_s$ stable,   & $\nu_s = \nu-1$ \\
&                          & $S_0>S_u>S_s$   &          \\[0.3ex]
\hline
& $\hat{\varepsilon} < 0:$ & $\xi_0$ stable,   & $\nu_0 = \nu+1$ 
\rule{0ex}{2.7ex} \\[1.0ex]
& $\hat{\varepsilon} > 0:$ & $\xi_0$ stable,   & $\nu_0 = \nu-1$ \\
{\large $\beta = -1$}
&                          & $\xi_u$ unstable, & $\nu_u = \nu$   \\
&                          & $\xi_s$ stable,   & $\nu_s = \nu+1$ \\
&                          & $S_s>S_u>S_0$   &          \\[0.3ex]
\hline
\end{tabular}
\end{center}
\caption{Properties of orbits which participate in a generic
period $m$-tupling bifurcation with $m \geq 5$. $\xi_0$ denotes
the central orbit, and $\xi_u$ and $\xi_s$ denote the unstable 
and stable satellite orbits, respectively.}
\end{table}

The bifurcations in the oval billiard that lead to the pairs
$e$, $h$ and $m$ are not generic. This is due to the symmetries
of the billiard system. But the normal forms which describe
these bifurcations agree with normal forms of generic
bifurcations, namely with those for double the repetition 
number. For example, the pair $e$ results from a bifurcation
of the 3-fold repetition of the orbit $a_s$, but its normal form
corresponds to that of a generic bifurcation with repetition
number $m=6$. This follows from the treatment of bifurcations
in the presence of symmetries \cite{Rim82,AMBD87}. It can be 
understood by considering a Poincar\'e section of surface
perpendicular to the central orbit. For a generic
period-$m$-tupling bifurcation with $m>4$, the map from the
Poincar\'e section of surface at some starting point to the
Poincar\'e section of surface after $m$ traversals of the
central orbit has $2 m$ fixed points near the central orbit
(after the bifurcation); $m$ of them correspond to the new
stable orbit and $m$ to the new unstable orbit, since both
orbits cross the Poincar\'e section of surface $m$ times
before they close. For the considered bifurcations in the
oval billiard there are $4 m$ instead of $2 m$ fixed points
in the vicinity of the central orbit, since two new stable
and two new unstable orbits arise, but the number
and arrangement of stable and unstable fixed points is
the same as for a generic period-$(2m)$-tupling
bifurcation.

As a consequence, also the period-tripling and
period-quadrupling bifurcations of the stable vertical
orbit $a_s$ in the oval billiard can be described by the
formula for generic bifurcations with $m>4$. The only
difference is, that $m$ has to be replaced by $2 m$
in Eq.\,(\ref{bif1}) and the whole formula has to
be multiplied by two, if one considers the bifurcation
in the full oval.

In the desymmetrized billiard, the quarter oval, the
orbit $a_s$ runs along a part of the boundary of the
billiard and Eq.\ (\ref{bif1}) has to be modified
in a different way. We describe this by first looking
at the Gutzwiller contribution of the $n$-th repetition
of the orbit $a_s$ which is modified to take account
of the fact that $a_s$ is a boundary orbit
\begin{eqnarray} \label{bif2}
\hat{d}_{na_s}(E) &=&
\frac{1}{\pi \hbar} \frac{T_1}{2 n \sqrt{|\Tr M_1 - 2|}}
\cos(\frac{S_1}{\hbar} - \frac{\pi}{2} \nu_1) + 
\frac{1}{\pi \hbar} \frac{T_2}{2 n \sqrt{|\Tr M_2 - 2|}}
\cos(\frac{S_2}{\hbar} - \frac{\pi}{2} \nu_2)
\nonumber \\ &=&
\frac{b}{2 \pi k} \left[ 
\frac{\sin(2 n k b)}{2 \sin(n v)} - 
\frac{\cos(2 n k b)}{2 \cos(n v)} \right] \, ,
\end{eqnarray}
where $b=1-\delta/3$ and 
$v = \arcsin(\sqrt{(1-\delta/3)/(1+\delta)})$. The second
line in Eq.\ (\ref{bif2}) is given in dimensionless units.
The quantities
with index 1 are those of the vertical stable orbit in the half
oval with Dirichlet boundary conditions on the
$x$-axis. Furthermore, $T_2=T_1$, $S_2=S_1$, $M_2=-M_1$ and
$\nu_2 = \nu_1 + 1 + 2([n v/\pi]\mbox{mod}2)$. 
Eq.\ (\ref{bif2}) has the form of a
sum of contributions from two orbits, each with half
the usual amplitude. Bifurcations of the $n$-th repetition
of the orbit $a_s$ occur
when $\delta \neq 0$ and one of the two terms in
(\ref{bif2}) diverges. The approximation (\ref{bif2})
is only valid if the orbit
is well separated from a bifurcation. Near a
bifurcation it has to be modified by replacing the
term which diverges at the bifurcation by Eq.\ 
(\ref{bif1}) where $m$ is twice the lowest repetition
number for which the bifurcation occurs and $r=2n/m$.
If both terms are close to a bifurcation then
both terms have to be replaced.

In the following we apply the uniform approximation
(\ref{bif1}) and its modifications in order to describe
the break-up of tori of the ellipse with $\alpha < 0$. 
A comparison with the semiclassical
contributions of the tori in the elliptical billiard
(section \ref{secnega}) shows that this
effectively corresponds to an application of the
broken torus approximation (\ref{bt3}) to the
torus term in Eq.\ (\ref{sc18}). The interference
term and the contribution of the bifurcating stable
orbit do not change their form. This is related to
the fact, that for a generic period-$m$-tupling 
bifurcation with $m>4$ the differences of the actions
of the satellite orbits grow more slowly than the
differences between the actions of the satellite
orbits and the action of the central stable orbit.
This is not the case for all bifurcations. For example,
there is a period-doubling bifurcation in the ellipse
for a ratio of the half-axis $a/b = \sqrt{2}$.
This bifurcation is again described by the
approximation Eq.\ (\ref{sc18}) (plus the contribution
of the stable orbit),
but in this case it is not correct to apply
the broken torus approximation (\ref{bt3}) to the
torus term in Eq.\ (\ref{sc18}) in order
to describe the break up of the torus. 
Instead one has to use the uniform approximation for
a generic bifurcation with $m=4$ which has a
more complicated form and is described by
a diffraction catastrophe integral for the 
catastrophe $X_9$ \cite{OH87}.

\subsection{Numerical results}

In the following we examine different approximations
for the spectral density numerically. For that purpose
the energy levels of the quarter oval with $\delta = 0.3$
were determined by a boundary element method up to
energy $E_{max}=20\,000$.
This corresponds to 1190 different energy levels.

In order to compare semiclassical results with
quantum mechanical results we consider
again the Fourier cosine transform $D(x)$ with a 
cut-off that is defined in (\ref{num1}). The damping
factor is again $z=5$. 
Figure~7 compares determinations
of $D(x)$ from the quantum mechanical spectrum 
$D_{\scriptscriptstyle QM}(x)$ and the uniform
approximation $D_{\scriptscriptstyle CUA}(x)$.
For the calculation of $D_{\scriptscriptstyle CUA}(x)$
all periodic orbits up to length $l=12$ and up to
30 reflections on the boundary were included.
Both curves are in good agreement and it is difficult
to see the differences. For that reason we plot
in the following figures the semiclassical error,
i.\,e.\ the difference between the quantum mechanical
curve $D_{\scriptscriptstyle QM}(x)$ and different
semiclassical (or uniform) approximations for it. 
\begin{figure}[p]
\begin{center}
\mbox{\epsfig{file=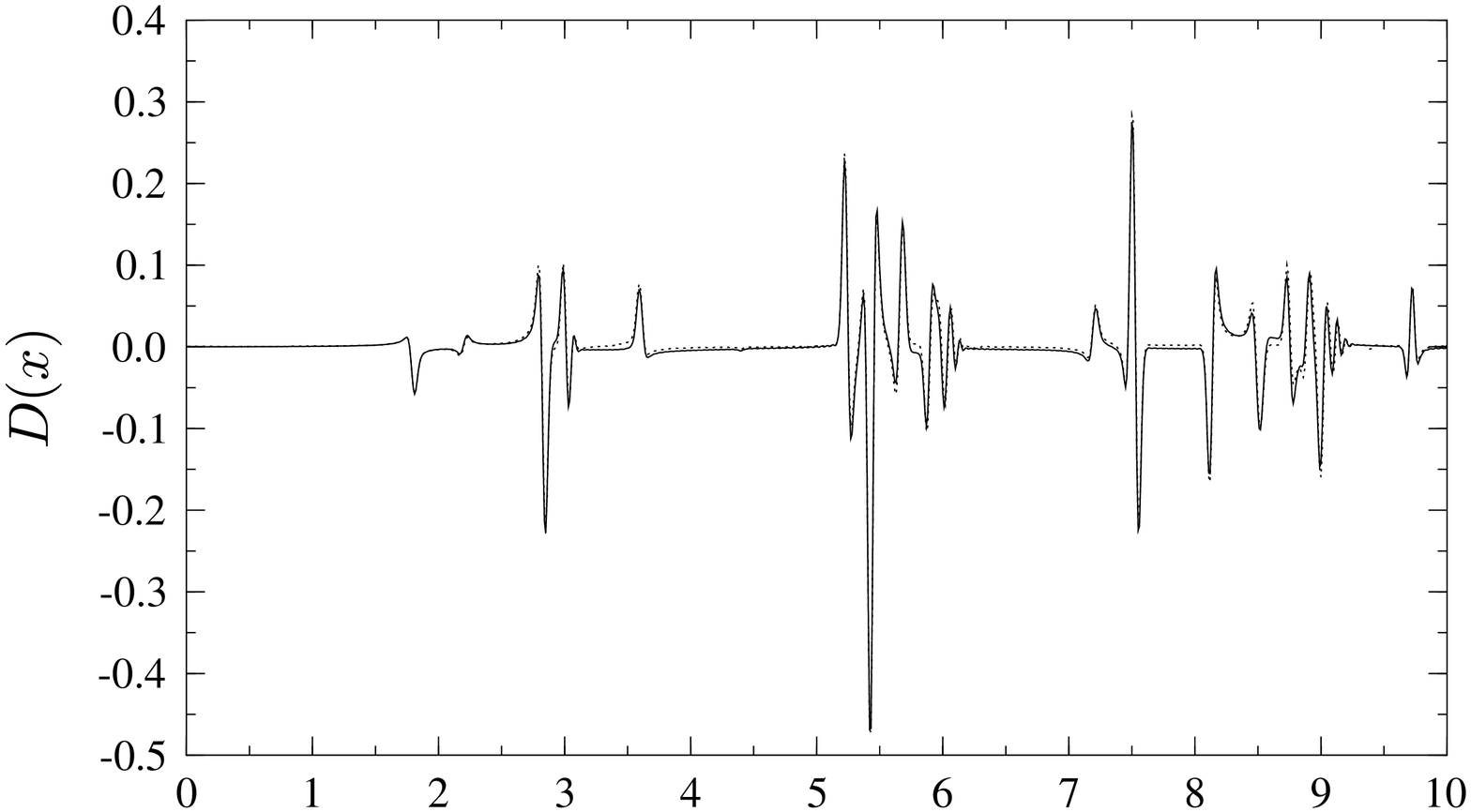,width=16cm,
bbllx=0pt, bblly=-21pt, bburx=750pt, bbury=380pt}}
\end{center}
\caption{The Fourier cosine transforms $
D_{\scriptscriptstyle QM}(x)$ (full) 
and $D_{\scriptscriptstyle CUA}$ (dashed) of the level 
density, calculated from
quantum energies and periodic orbits, respectively.}
\vspace*{-0.2cm}
\begin{center}
\mbox{\epsfig{file=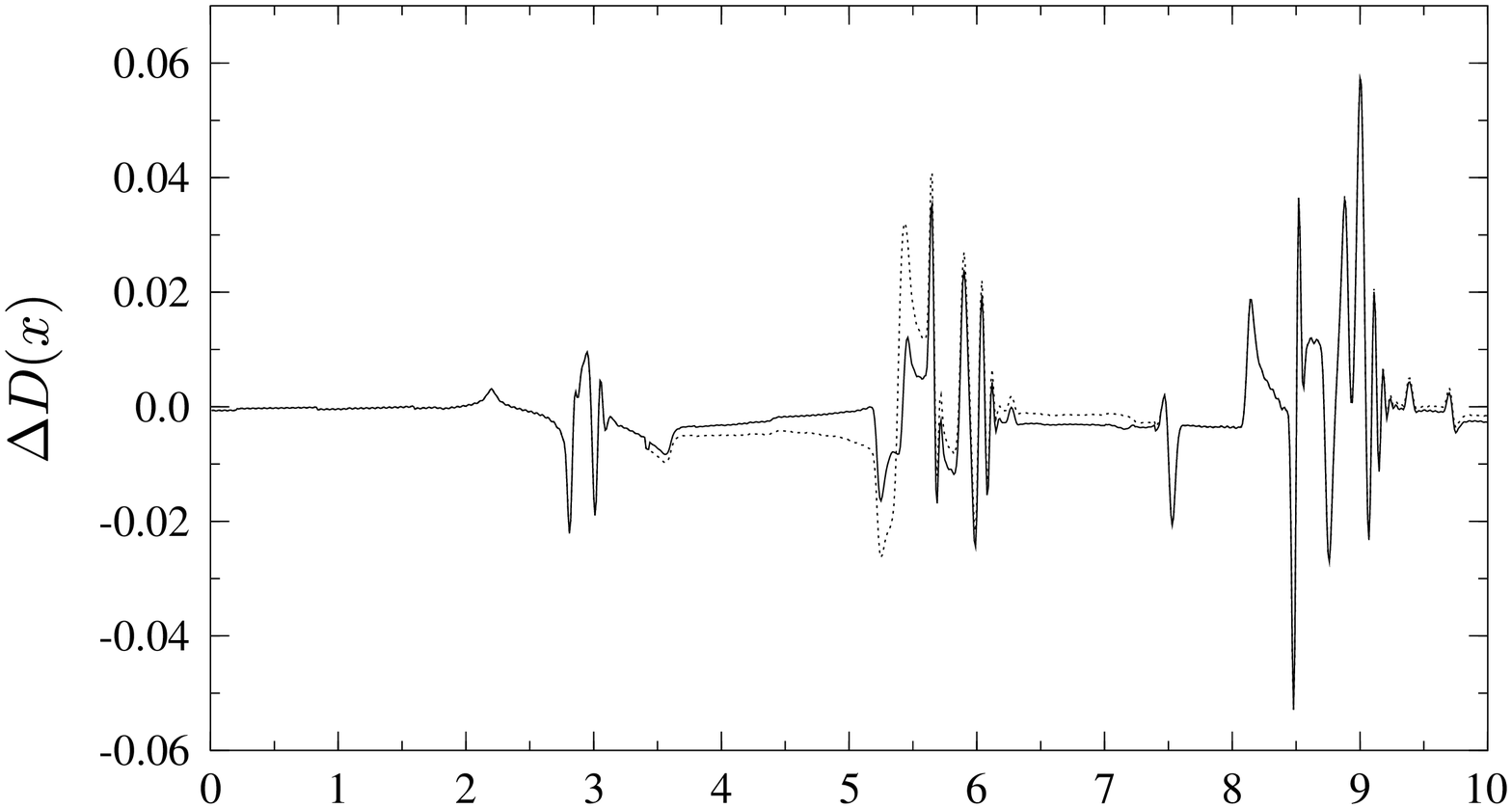,width=16cm,
bbllx=0pt, bblly=-21pt, bburx=750pt, bbury=380pt}}
\end{center}
\caption{The semiclassical errors $D_{\scriptscriptstyle QM}(x) 
- D_{\scriptscriptstyle QT}(x)$ (dotted)
and $D_{\scriptscriptstyle QM}(x) - D_{\scriptscriptstyle QTI}(x)$
(full) for the quasi-torus approximation
with and without interference terms, respectively.}
\end{figure}

The most basic semiclassical approximation for a
quasi-integrable system, i.\,e.\ for a slightly
perturbed integrable system, is the quasi-torus
approximation. In this approximation the pairs
of periodic orbits are treated as if they
still contribute like a torus of periodic orbits.
We use here a quasi-torus approximation in which
the semiclassical amplitudes and phases are
determined directly from the corresponding
periodic orbits. This is done by setting 
$\Delta S$ equal to zero in Eq.\ (\ref{bt3}). 
This approximation is used for all pairs of orbits
except for the pair $a$ and its repetitions. 
For this pair the action difference is already quite
large, and the orbits are treated as separate periodic
orbits. In more detail, the contribution
of the $n$-th repetition of the unstable orbit $a_u$
is approximated by
\begin{equation}
\hat{d}_{n a_u}(E) = \frac{a (-1)^n}{2 \pi k} \left[
\frac{\cos(2 n k a - n \frac{\pi}{2})}{2 \sinh(n u)} - 
\frac{\cos(2 n k a - n \frac{\pi}{2})}{2 \cosh(n u)} 
\right] \, ,
\end{equation}
with $a=1+\delta/3$ and 
$u = \mbox{arccosh}(\sqrt{(1+\delta/3)/(1-\delta)})$,
and the contributions of the stable orbit $a_s$ and
its repetitions by Eq.\ (\ref{bif2}).

The semiclassical error for the quasi-torus approximation 
$D_{\scriptscriptstyle QM}(x) - D_{\scriptscriptstyle QT}(x)$
is plotted as dotted line
in figure~8. This approximation already works
relatively good. This is due to the fact
that the action differences within pairs of orbits
are still relatively small and that the orbits
are not very close to a bifurcation.
However, in comparison with the semiclassical error for
the ellipse in figure~3 the semiclassical error for the
quasi-torus approximation in figure~8 is clearly
larger. The scales of the plots differ by approximately 50\%.
In order to improve the approximation we apply Eq.\
(\ref{bif1}) for the pairs of orbits which arose from
a bifurcation, i.\,e.\ for the pairs $e$, $h$ and $m$,
but we still set the action differences in formulas
(\ref{bt3}) and (\ref{bif1}) equal to zero. Note that
with this substitution Eq.\ (\ref{bif1}) has exactly
the same form as that of the contribution
of a torus near a bifurcation (see section \ref{secnega}). 
This approximation can thus be
considered as an improved quasi-torus approximation which
takes into account cases in which quasi-tori are close to
a bifurcation. The corresponding semiclassical error
$D_{\scriptscriptstyle QM}(x) - D_{\scriptscriptstyle QTI}(x)$
is plotted in figure~8
as full line. One can see an clear improvement near
$x \approx 5.4$. This length corresponds to the pair
$e$ which is that pair out of $e$, $h$ and
$m$ which is closest to a bifurcation. This can be seen
from the differences between the mean lengths of the pairs
and the $m$-th repetition of the central orbit from
which they arose by a bifurcation:
$\bar{l}_e - 3 l_{a_s} = 0.03636$, 
$\bar{l}_h - 4 l_{a_s} = 0.34064$, and
$\bar{l}_m - 5 l_{a_s} = 0.72132$. The pairs $h$ and $m$
are already well separated from the orbit $a_s$ and
thus the inclusion of the interference term does not
improve the approximation.

For a further improvement of the semiclassical
approximation
we apply now the uniform approximations (\ref{bt3})
and (\ref{bif1}) without setting $\Delta S$ equal to
zero. The result $D_{\scriptscriptstyle QM}(x) - 
D_{\scriptscriptstyle UA}(x)$ is shown in
figure~9 as full line. For a comparison with the
previous approximation the function 
$D_{\scriptscriptstyle QM}(x) - D_{\scriptscriptstyle QTI}(x)$
is plotted again as dotted line. One can observe a
clear improvement of the approximation at three places
which correspond to the first three repetitions of the
pair $b$. Table 1 shows that this is the pair
for which the length difference is largest,
and the size of the semiclassical error increases with the
size of $\Delta l$. We note that for the decrease of
the semiclassical error it is important to use the
whole formula (\ref{bt3}) and not only the first term
with the $\mbox{J}_0$-Bessel function. A criterion
for the importance of the second term is the size
of $\Delta t$ in Table 1 which increases
rapidly with the repetition number of the pair $b$.
\begin{figure}[thbp] 
\begin{center}
\mbox{\epsfig{file=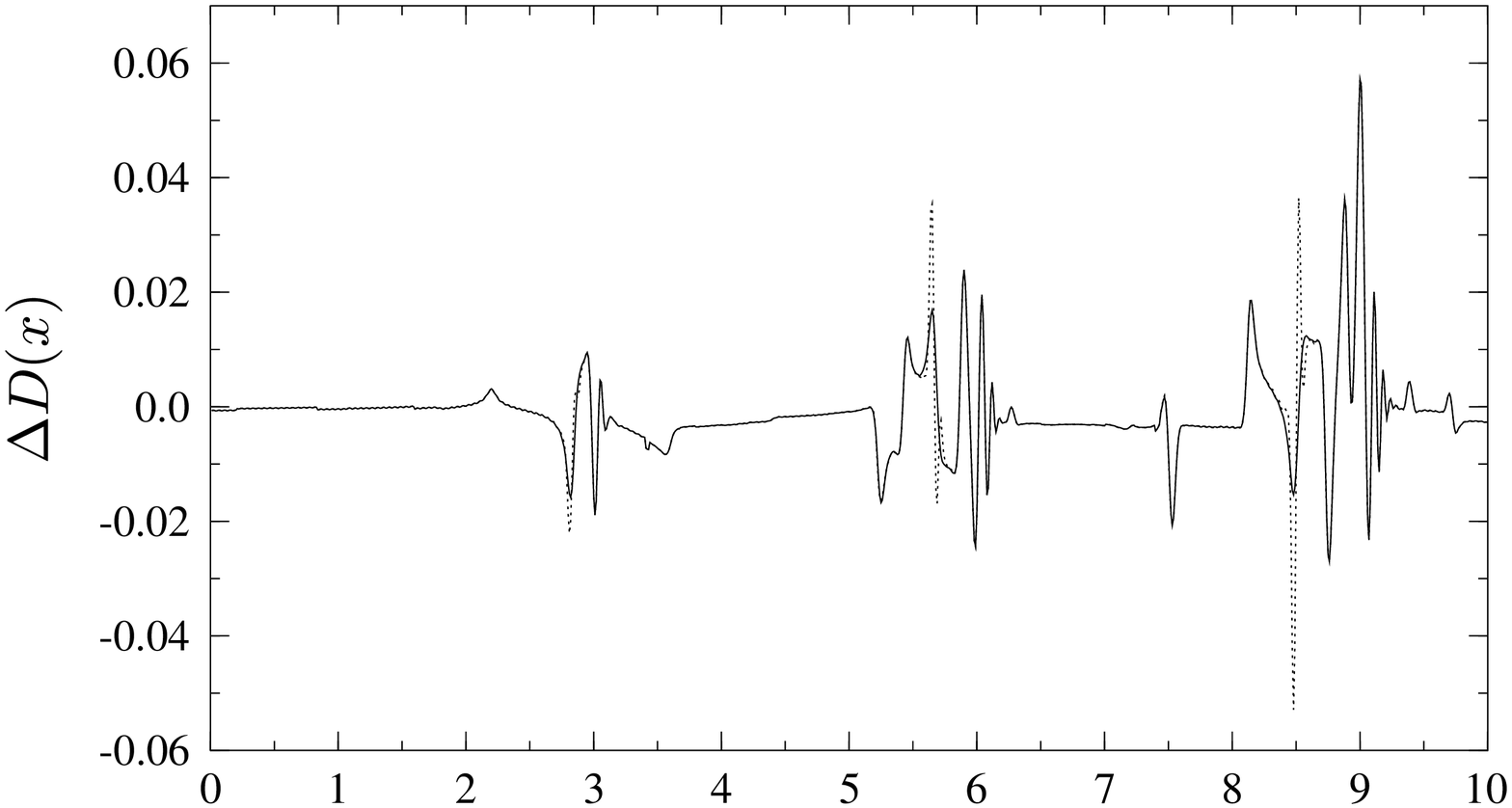,width=16cm,
bbllx=0pt, bblly=-21pt, bburx=750pt, bbury=380pt}}
\end{center}
\caption{The semiclassical errors $D_{\scriptscriptstyle QM}(x) 
- D_{\scriptscriptstyle QTI}(x)$ (dotted)
and $D_{\scriptscriptstyle QM}(x) - D_{\scriptscriptstyle UA}(x)$
(full) for the quasi-torus approximation
with interference term and for the uniform approximation,
respectively.}
\vspace*{-0.2cm}
\begin{center}
\mbox{\epsfig{file=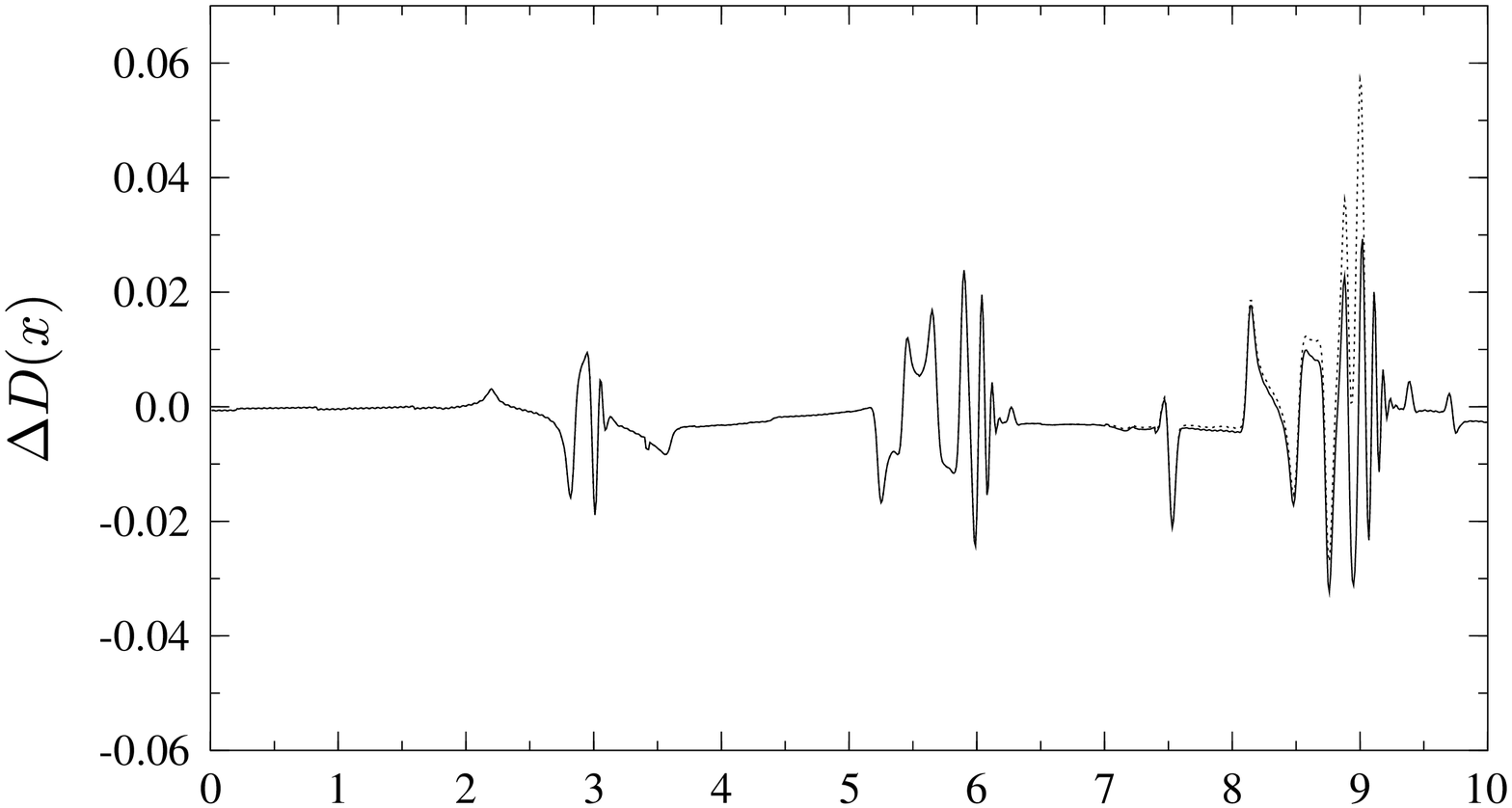,width=16cm,
bbllx=0pt, bblly=-21pt, bburx=750pt, bbury=380pt}}
\end{center}
\caption{The semiclassical errors $D_{\scriptscriptstyle QM}(x) 
- D_{\scriptscriptstyle UA}(x)$ (dotted)
and $D_{\scriptscriptstyle QM}(x) - D_{\scriptscriptstyle CUA}(x)$
(full) for the uniform approximation
with and without the contribution of a pair of complex orbits,
respectively.}
\end{figure}

Even for the uniform approximation there is still a relatively
large error at $x \approx 9$. This error is due to the fact
that up to now we considered only contributions from real
orbits to the spectral density. But before a bifurcation
there are also contributions from complex orbits as is
expressed by the interference term in Eq.\ (\ref{bif1}). 
The error at $x \approx 9$ is due to
fact that the 5-fold repetition of the orbit $a_s$ is close
to a second bifurcation which occurs at 
$\delta_{bif} = (75-24\sqrt{5})/61 \approx 0.3497$.
In a last step we improve the approximation by taking
into account the contribution of these complex orbits.
This is done by determining these orbits for values
$\delta > \delta_{bif}$ where they are real. From these data
the properties of the orbits can be extrapolated
into the complex region. The final result
$D_{\scriptscriptstyle QM}(x) - D_{\scriptscriptstyle CUA}(x)$
is shown in figure~10
as full line. For a comparison with the previous
approximation the function $D_{\scriptscriptstyle QM}(x) - 
D_{\scriptscriptstyle UA}(x)$
is plotted again as dotted line. As one can see,
the error is clearly decreased near $x=9$ by
including the contribution of the complex orbits,
and the final semiclassical error is of the same
size as in the elliptical billiard.
This shows that by using the uniform approximations
(\ref{bt3}) and (\ref{bif1}) the same accuracy of 
the semiclassical approximation can be achieved
as in the unperturbed integrable system.

\section{Conclusions}

Most applications of semiclassical trace formulas in
the literature have concentrated on cases where the
periodic orbits are either isolated or occur in
families. These approximation are valid only
in restricted classes of systems.
In general systems, it is common that
semiclassical contributions of periodic orbits
cannot be considered isolated from contributions
of other periodic orbits in their neighbourhood.
This requires the use of uniform approximations
which take account of the underlying classical
structures and yield
joint contributions of neighbouring periodic orbits.
In systems where the classical phase space has
regular and chaotic regions this is a generic situation,
and semiclassical quantization rules in terms of periodic
orbits in these systems always involve uniform approximations. 
But uniform approximations can also be necessary 
in integrable or chaotic systems, for example when
bifurcations of periodic orbits occur in integrable
systems or when periodic orbits are close to creeping or
diffractive orbits in chaotic billiard systems
\cite{PSSU96,SPS96}.

In the first part of this paper we examined an integrable
system, a billiard in form of an ellipse, and derived a
trace formula for it. The elliptical billiard is a standard
example for an integrable system, yet its semiclassical
trace formula is more complicated than a summation
over semiclassical contributions of isolated tori.
These complications are due to the presence of the
separatrix and of the stable periodic orbit. 
We were particularly interested in the treatment of the
bifurcations of the stable orbit and its connection to
generic bifurcations when the ellipse is perturbed. 
We derived a semiclassical trace formula which takes
account of these bifurcations, and we examined its
semiclassical accuracy. 

In the second part of this paper we deformed the
ellipse into an oval which is non-integrable. It
was shown that in the quasi-integrable regime
it is not always sufficient to treat the break-up
of tori by the general formulas of 
\cite{Ozo86,TGU95,UGT96}.
If a torus in the integrable system is close to a
bifurcation, then uniform formulas for general 
bifurcations are needed for describing its break-up.
In the numerical section it was shown
that with the uniform formulas
for the break-up of tori and for the bifurcations
the same semiclassical accuracy can be achieved
as in the unperturbed integrable system.

If the billiard system is deformed more, then the
stable periodic orbits of the billiard undergo further
bifurcations. A semiclassical description of a
system in the truly mixed regime, i.\,e.\ not
in the near-integrable regime, requires the use of
uniform approximations for all generic bifurcations 
\cite{SS96}. In case that the system has discrete
symmetries there are further bifurcations
which are specific for the considered symmetry. 
The application of uniform approximations for these
bifurcations requires
not only the determination of all real periodic orbits
up to some length, but also of all complex periodic
orbits which are close to becoming real. Additional
complications can arise when a periodic orbit undergoes
several subsequent bifurcations which cannot be considered
isolated. This shows, as is well-known, that
semiclassical trace formulas for systems with mixed
phase space are definitely more complicated as for
integrable or chaotic systems.

\bigskip \bigskip \noindent {\large \bf Acknowledgments} 
\vspace{2.5mm}

I am obliged to E.\ Bogomolny for many helpful suggestions
and discussions. It is a pleasure to thank also O.\ Bohigas,
M.\ Brack and H.\ R.\ Dullin for fruitful discussions, and,
furthermore, H.\ R.\ Dullin for communicating results of 
\cite{DWW96} before publication. 
This work was supported by the Alexander von Humboldt-Stiftung
and by the Deutsche Forschungsgemeinschaft under contract 
No.\ DFG-Ste 241/7-1 and /6-1.

\bibliographystyle{unsrt}
\bibliography{elpaper}

\end{document}